\documentclass[twocolumn]{autart}
\usepackage[T1]{fontenc}
\usepackage{graphicx}
\usepackage{xpatch}
\usepackage[noadjust]{cite}
\usepackage{amsfonts, mathtools, mathrsfs}
\usepackage{amssymb,mathrsfs, mathtools, float}
\DeclarePairedDelimiter{\norm}{\lVert}{\rVert}

\usepackage{xcolor}
\usepackage{tikz}
\usepackage{pgfplots}
\usetikzlibrary{matrix,decorations.pathreplacing,calc}
\newtheorem{assumption}{Assumption}
\newtheorem{definition}{Definition}
\usepackage{comment}
\usepackage{eso-pic}
\usepackage[
bookmarks=false,
colorlinks=false,
hyperfootnotes=false,
pdfborder={0 0 1}
]{hyperref}

\begin{document}
	
\AddToShipoutPictureFG*{%
	\begin{tikzpicture}[remember picture, overlay]
		\node[
		anchor=south,
		draw=white,
		fill=white,
		text=black,
		line width=0.5pt,
		minimum width=\linewidth,%0.8\paperwidth,
		minimum height=12mm,
		align=center
		] at ([xshift=30mm, yshift=25mm]current page.south)
		{};
		
		\node[
		anchor=south,
		draw=white,
		fill=white,
		text=black,
		line width=0.5pt,
		minimum width=\linewidth,%0.8\paperwidth,
		minimum height=12mm,
		align=center
		] at ([xshift=150mm, yshift=25mm]current page.south)
		{};
		
		\node[
		anchor=south,
		draw=white,
		fill=white,
		text=black,
		line width=0.5pt,
		minimum width=\linewidth,%0.8\paperwidth,
		minimum height=12mm,
		align=center
		] at ([xshift=105mm, yshift=25mm]current page.south)
		{This is the version of the original paper published in Automatica.  \href{https://doi.org/10.1016/j.automatica.2026.113108}{DOI: 10.1016/j.automatica.2026.113108}.};
	\end{tikzpicture}%
}

\begin{frontmatter}
\title{Robust and efficient data-driven predictive control}

\thanks[footnoteinfo]{M. Alsalti and M. Barkey contributed equally to this work.\\ Parts of this paper have been presented at the 2024 European Control Conference, June 25-28, 2024, Stockholm, Sweden.\\ Corresponding author: Mohammad Alsalti. \\ \textit{Email Addresses}: \texttt{\{alsalti,barkey,lopez,mueller\}\\@irt.uni-hannover.de}}

\author{Mohammad Alsalti$^\star$},
\author{Manuel Barkey$^\star$},
\author{Victor G. Lopez},
\author{Matthias A. M\"uller}

\address{Leibniz University Hannover, Institute of Automatic Control, 30167 Hannover, Germany}

\begin{abstract}
	We propose a robust and efficient data-driven predictive control (eDDPC) scheme which is more sample efficient (requires less offline data) compared to existing schemes, and is also computationally efficient. This scheme employs a recently proposed data-based representation of linear time-invariant (LTI) systems as a predictor. Such a representation serves as an alternative to Hankel-based predictors obtained from, e.g., the so-called \textit{fundamental lemma}, and can be derived by exploiting the kernel structure of shallow Hankel matrices of data. This allows for application of our proposed scheme using very short (and potentially irregularly measured) noisy input-output data, the amount of which is independent of the prediction horizon. To account for measurement noise, we provide a novel result that quantifies the uncertainty between the true (unknown) restricted behavior of the system and the estimated one from noisy data. Furthermore, we show that the robust eDDPC scheme is recursively feasible and that the resulting closed-loop system is practically exponentially stable. Finally, we compare the performance of this scheme to existing ones on a case study of a four tank system.
\end{abstract}
\vspace{-0.5em}
\begin{keyword}
Robust data-driven predictive control, behavioral approach, uncertainty quantification, SVD perturbations.
\end{keyword}

\end{frontmatter}

\section{Introduction}\label{sec_introduction}
\vspace{-0.5em}
Model predictive control (MPC) \cite{Rawlings17} is a powerful optimization-based control technique that is applicable to multivariable linear and nonlinear systems. MPC uses a model of the system being controlled in order to predict the behavior of the system over a finite horizon. In contrast, direct data-driven predictive control (DDPC) schemes have recently been developed using results from behavioral systems theory \cite{Willems86}. There, non-parametric representations given by the image of data matrices (see, e.g., \cite{Willems05, Markovsky22}) are employed as predictive ``models'' (cf. \cite{Coulson20}). Open-loop robustness and closed-loop guarantees of DDPC schemes for LTI systems were established in \cite{Coulson22, Berberich203, Berberich22_IRP}. Many extensions soon followed including nonlinear \cite{Verhoek21, Lian21, Berberich204, Alsalti2021c, lazar2024basis}, stochastic \cite{Pan22, Breschi23} and distributed DDPC \cite{Kohler22,Alonso22} among others. To account for noisy data, different regularization techniques were proposed (see, e.g., \cite{Doerfler23, Breschi23_autoDDC} and the references therein). DDPC schemes have also been successfully applied to various real-world systems, e.g., power systems \cite{markovsky2023powersys} and quadcopters \cite{elokda2021data}, thus making DDPC an important and well-established control technique. We refer to \cite{verheijen2023handbook, berberich2024overview} for comprehensive surveys of DDPC and its extensions.

Successful application of DDPC schemes requires that the collected offline data is sufficiently rich. This is typically ensured by imposing suitable persistence of excitation (PE) conditions on the input. However, PE necessitates that the data is sufficiently long and its length increases with increased system order, number of inputs and prediction horizon length. As a result, the computational complexity of solving the corresponding optimal control problem increases. Existing works that address efficiency in DDPC either focus on sample efficiency (using less data) by, e.g., segmentation of the prediction horizon \cite{Dwyer23}, or on computational efficiency by reducing the number of decision variables through the use of, e.g., singular value decomposition of the data matrices \cite{zhang2022}. None of the above schemes, however, simultaneously addresses sample and computational efficiency of DDPC schemes. In fact, the segmentation procedure in \cite{Dwyer23} results in an increased number of decision variables, whereas \cite{zhang2022} requires the availability of sufficiently long data. In some practical applications, it may not be easy to obtain long PE data that allow for application of DDPC schemes with long prediction horizons. Moreover, data can be irregularly measured due to sensor failure or inability to measure data consecutively (as in, e.g., biomedical applications). Recently, an alternative non-parametric representation of the finite-length behavior of LTI systems was proposed in \cite{Alsalti2023_md} using a (potentially short and irregularly measured) data sequence. In this work, we will exploit the results of \cite{Alsalti2023_md} to simultaneously address sample and computational efficiency in DDPC.

\emph{Contributions:} First, we propose a sample- and computationally-efficient robust data-driven predictive control (eDDPC). Unlike existing methods, this scheme can also potentially be employed when only short and irregularly measured (noisy) offline data is available. In the preliminary conference version of this work (see \cite{alsalti2023eddpc_ecc}), we presented the nominal noise-free setting. To account for measurement noise, we provide as a second contribution a novel result on uncertainty quantification in the behavioral framework. In particular, we derive a bound on the angles between two subspaces: the unknown finite-length behavior of the system and its known approximation. Our results rely on investigating SVD perturbations of the data matrices (cf. \cite{Stewart91}). In contrast, existing works on uncertainty quantification in the behavioral framework \cite{Padoan22, Fazzi23} measure the distance between two \textit{known} behaviors. As a third contribution, we show that the robust eDDPC scheme is recursively feasible and that the closed-loop system is practically exponentially stable in the presence of input constraints. This is different from existing efficient DDPC schemes (e.g., \cite{Dwyer23,zhang2022}) where no theoretical guarantees were provided. Finally, we analytically and numerically compare the performance of this scheme to existing ones on a case study of a four tank system.

Section~\ref{sec_preliminaries} introduces necessary background material. Section~\ref{sec_eDDPC} presents the nominal eDDPC scheme. Section~\ref{sec_uncertainty_quantification} includes a novel result on uncertainty quantification in the behavioral framework. Section~\ref{sec_rob_eDDPC} formulates the robust eDDPC scheme in presence of noise and establishes stability guarantees. Section~\ref{sec_examples} includes a simulation case study and Section~\ref{sec_conclusions} concludes the paper.

\emph{Notation:} The sets of integers, natural and real numbers are denoted by $\mathbb{Z},\mathbb{N},\mathbb{R}$, respectively. The restriction of integers is denoted by $\mathbb{Z}_{[a,b]}$ for $b>a\in\mathbb{Z}$. For a matrix $M\in\mathbb{R}^{m\times n}$, we denote its image by $\mathrm{im}(M)$ and its kernel by $\mathrm{ker}(M)$. We use $\mathrm{null}(M)$ to denote an operator which returns a basis of $\mathrm{ker}(M)$. The singular values of $M$ are ordered scalars $s_1(M)\geq \cdots \geq s_{\min\{m,n\}}(M)\geq0$. We use $\norm*{M}_i$, $i\in\{2,\infty,F\}$, to denote the induced norms or the Frobenius norm, respectively. We write $\textup{diag}(M_1,M_2)$ to denote the block-diagonal concatenation of matrices $M_1,M_2$. The largest (respectively, smallest) eigenvalues of a symmetric positive definite matrix $P\succ0$ are denoted by $\lambda_{\textup{max}}(P)$ ($\lambda_{\textup{min}}(P)$). The weighted norm of a vector $x$ is $\norm*{x}_P\coloneqq\sqrt{x^\top Px}$, whereas $\norm*{x}_i$, $i\in\{1,2,\infty\}$ denotes the standard vector norms. A function $\phi:\mathbb{R}_{\geq0}\to\mathbb{R}_{\geq0}$ is of class $\mathcal{K}_{\infty}$ if it is continuous, zero at zero, strictly increasing and $\lim_{r\to\infty}\phi(r)=\infty$. For $T\in\mathbb{N}$, the set of finite-length $q-$variate time series $w=(w_0,\ldots,w_{T-1})$ is denoted by $\left(\mathbb{R}^{q}\right)^T$. We also use $w$ to denote the stacked vector $w=[\begin{matrix}w_0^\top & \cdots & w_{T-1}^\top\end{matrix}]^\top\in~\mathbb{R}^{qT}$, and a window of it by $w_{[a,b]}$ where $0\leq a < b \leq T-1$. The Hankel matrix of depth $L\leq T$ of $w$ is defined as\linebreak $\mathscr{H}_L(w) \coloneqq \begin{bmatrix}w_{[0,L-1]} & w_{[1,L]} & \cdots & w_{[T-L,T-1]}\end{bmatrix}$.
\vspace{-1em}
\section{Preliminaries}\label{sec_preliminaries}
\vspace{-0.5em}
\subsection{SVD perturbations}\label{sec_SVDperturbations}
\vspace{-0.5em}
Consider a matrix $M\in\mathbb{R}^{m\times n}$ with $m\leq n$ and let its perturbation be $\widehat{M} = M+E$ with $\norm*{E}_F<\infty$, such that $0<r\coloneqq\mathrm{rank}(M)\leq \rho\coloneqq\mathrm{rank}(\widehat{M}) \leq m$. We consider a decomposition of $M$ and $\widehat{M}$ of the following form
\vspace{-0.5em}
\begin{equation}
	\begin{aligned}
		M &= \big[\begin{matrix}U_M & W_M\end{matrix}\big]\textup{diag}(S_M,0)\big[\begin{matrix} V_M & Q_M \end{matrix} \big]^\top,\\
		\widehat{M} &= \big[\begin{matrix}U_{\widehat M} & W_{\widehat M}\end{matrix}\big]\textup{diag}(S_{1,\widehat M},S_{2,\widehat M})\big[\begin{matrix}V_{\widehat M} & Q_{\widehat M}\end{matrix}\big]^\top,
	\end{aligned}\label{eqn_svdofMs}
	\vspace{-0.5em}
\end{equation}
where $S_M=\textup{diag}(s_1(M),\ldots,s_r(M))$ (similarly for $S_{1,\widehat{M}}$), whereas $S_{2,\widehat{M}}$ contains the remaining $m-r$ singular values of $\widehat{M}$. The matrices $U_{M},W_M$ and $V_M,Q_M$ (similarly for $U_{\widehat M},W_{\widehat M},V_{\widehat M},Q_{\widehat M}$) are semi-orthonormal matrices of appropriate dimensions whose columns represent the right and left singular vectors, respectively. The image of these matrices are known as the \textit{singular subspaces}. The following theorem states that the singular values of perturbed matrix $\widehat M$ remain in a neighborhood around the corresponding singular values of $M$.
\vspace{-0.5em}
\begin{lem}\textup{\cite{Stewart91}}\label{thm_Mirsky}
	Given $M,\widehat{M}= M+E \in \mathbb{R}^{m\times n}$ where $0<r=\mathrm{rank}(M)\leq \rho=\mathrm{rank}(\widehat{M}) \leq m$ and $\norm*{E}_F<\infty$, let their SVDs be given by \eqref{eqn_svdofMs}. Then,
	\vspace{-0.5em}
	\begin{equation}
		\sqrt{\sum\nolimits_{i=1}^{\rho} \left(s_i(\widehat{M}) - {s}_i(M)\right)^2} \leq \norm*{E}_F.\label{eqn_Mirsky}
		\vspace{-0.5em}
	\end{equation} 
\end{lem}%
\vspace{-1em}
The effect of the disturbance $E$ on the singular subspaces is analyzed using the \textit{principal angles} between the true and perturbed subspaces, which are defined as follows.
\begin{definition}\textup{\cite{Miao1992}}\label{def_principalangles}
	Let $\mathcal{X},\,\mathcal{Y}$ be two subspaces of $\mathbb{R}^n$ of dimension $k\leq n$. The principal angles $\theta_i$, $0\leq\theta_i\leq\pi/2$ for $i\in\mathbb{Z}_{[1,k]}$, between $\mathcal{X},\,\mathcal{Y}$ are recursively defined by
	\vspace{-1em}
	\begin{align}
		\cos\theta_i = \bar{x}_i^\top\bar{y}_i = &\arg\max\limits_{x_i\in\mathcal{X}}\max\limits_{y_i\in\mathcal{Y}}\quad x_i^\top y_i\notag\\
		&\textup{s.t. }\,\norm*{x_i}_2=\norm*{y_i}_2=1,\label{eqn_PA}\\
		&\qquad x_i^\top \bar{x}_{j}=0, \, y_i^\top \bar{y}_{j}=0, \, \forall j\in\mathbb{Z}_{[1,i-1]},\notag
	\end{align}
	where $\bar{x}_i,\bar{y}_i\in\mathbb{R}^n$, $i\in\mathbb{Z}_{[1,k]}$, are principal vectors which form orthonormal bases for $\mathcal{X,Y}$, respectively.
\end{definition}
\vspace{-0.75em}
Principal angles are unique and the principal vectors, although not unique, always exist. In the following, we denote the matrix of principal angles between $\mathcal{X,Y}$ as $\Theta(\mathcal{X},\mathcal{Y})=\mathrm{diag}(\theta_1,\ldots,\theta_{k})$. The following theorem states that the angles between the true and perturbed singular subspaces\footnote{Here, we recall the result only in terms of $\Theta(\mathrm{im}(U_M),\mathrm{im}(U_{\widehat{M}}))$ for notational simplicity, however the same bound holds for all other singular subspaces.} can be bounded in terms of $\norm*{E}_F$.
\vspace{-0.5em}
\begin{lem}\textup{\cite{Stewart91}}\label{thm_wedin}
	Consider the setting of Lemma~\ref{thm_Mirsky}. Then,
	\vspace{-1em}
	\begin{equation}
		\norm{\sin (\Theta(\mathrm{im}(U_M),\mathrm{im}(U_{\widehat{M}})))}_F \leq (\sqrt{2}/\delta)\norm*{E}_F.
		\label{eqn_Wedin}
		\vspace{-1em}
	\end{equation}
	where $\delta = s_r(\widehat{M})$, and $\sin(\cdot)$ is applied element-wise.
\end{lem}
\vspace{-0.5em}
In the following lemma, we show that if a basis for one singular subspace (e.g., $\mathrm{im}(U_{\widehat{M}})$) is fixed, then there exists a basis for the other (potentially unknown) subspace (e.g., $\mathrm{im}(U_{M})$) such that their difference is bounded by the angles between the two subspaces and, hence, also bounded by $\norm{E}_F$ as in Lemma~\ref{thm_wedin}. This is an important result that will later be used in Section~\ref{sec_uncertainty_quantification}.
\vspace{-0.5em}
\begin{lem}\label{lemma_newbasis}
	Consider $M,\widehat{M}= M+E \in \mathbb{R}^{m\times n}$ where $0<r=\mathrm{rank}(M)\leq \rho=\mathrm{rank}(\widehat{M}) \leq m$ and $\norm*{E}_F<\infty$, along with their decomposition \eqref{eqn_svdofMs}. Then, there exists $\widetilde{U}_{M}$ such that $\mathrm{im}(\widetilde{U}_{M})=\mathrm{im}(U_M)$ and
	\vspace{-0.75em}
	\begin{equation}
		\norm{U_{\widehat{M}} - \widetilde{U}_{M}}_F \leq 2\sqrt{r}\norm{\sin (\Theta(\mathrm{im}(U_M),\mathrm{im}(U_{\widehat{M}})))}_F\hspace{-0.5mm},\label{eqn_newbasis}
		\vspace{-0.75em}
	\end{equation}
	where $\delta = s_r(\widehat{M})$, and $\sin(\cdot)$ is applied element-wise.
\end{lem}\vspace{-0.5cm}
\begin{pf}
	See Appendix~\ref{appendix_newbasis}.
\end{pf}
\vspace{-0.5cm}

\subsection{Behavioral approach to systems theory}
\vspace{-0.5em}
In the behavioral approach to systems theory \cite{Willems86}, a system is viewed as a tuple $(\mathbb{N},\mathbb{R}^q,\mathscr{B})$, where $\mathbb{N}$ is the time axis, $\mathbb{R}^q$ is the signal space and $\mathscr{B}$ is the \textit{behavior} of the system. The finite-length behavior $\mathscr{B}|_T$ is defined as a set of finite-length trajectories. A trajectory of length $T$ of the system is denoted by $w\in\left.\mathscr{B}\right|_T$, where $w_t=\begin{bsmallmatrix} u_t \\ y_t \end{bsmallmatrix}$ is partitioned to inputs $u_t\in\mathbb{R}^m$ and outputs $y_t\in\mathbb{R}^p$. The set of discrete-time LTI systems with $q$ variables and known\footnote{When only an upper bound on $n$ is known, inferring the true system's order from data is possible under certain conditions, e.g., noise-free data or high signal-to-noise ratio. For simplicity, we consider systems of known complexity.} complexity $(m,n,\ell)$ is denoted by $\partial\mathscr{L}_{m,n,\ell}^{q}$, where $q=m+p$, $n$ is the order of the system and $\ell$ is the lag of the system (observability index), respectively.

A \textit{kernel representation} of $\mathscr{B}\in\partial\mathscr{L}_{m,n,\ell}^{q}$ is given by~\cite{Willems86}
\vspace{-0.75em}
\begin{equation}
	\mathscr{B} = \textup{ker} (R(\sigma)) = \{ w\in(\mathbb{R}^q)^{\mathbb{N}} ~|~ R(\sigma)w_t=0\},\label{eqn_kernelpolymatrix}
	\vspace{-0.5em}
\end{equation}
where $\sigma^j w(k) \coloneqq w(k+j)$, for $j\in\mathbb{N}$, is the shift operator and $R(\sigma)$ is a polynomial matrix. In \cite{Markovsky22} it was shown that $\mathrm{im}(\mathscr{H}_L(w)) = \mathscr{B}|_L$ if and only if
\vspace{-0.75em}
\begin{equation}
	\mathrm{rank}(\mathscr{H}_L(w)) = mL+n,\label{eqn_GPE}
	\vspace{-0.75em}
\end{equation}
for any $L\geq\ell$. When this holds, one obtains a data-based representation of all length-$L$ trajectories of the system, i.e., $\bar{w}\in\mathscr{B}|_L$ if and only if $\exists\alpha\in\mathbb{R}^{T-L+1}$ such that
\vspace{-0.75em}
\begin{equation}
	\mathscr{H}_L(w)\alpha = \bar{w}.\label{GPE_FL}
	\vspace{-0.75em}
\end{equation}
For controllable systems, persistence of excitation (PE, see Definition~\ref{def_PE} below) of the input ensures that \eqref{eqn_GPE} holds. This latter result is known as the \textit{fundamental lemma}, see \cite[Th. 1]{Willems05}. 
\begin{definition}\textup{\cite{Willems05}}\label{def_PE}
	A sequence $u\in(\mathbb{R}^m)^{T}$ is said to be persistently exciting of order $L$ if $\textup{rank}(\mathscr{H}_L(u))=mL$.
\end{definition}
Another result which follows from the rank condition~\eqref{eqn_GPE} is \textit{identifiability from data} which allows us to retrieve a kernel representation \eqref{eqn_kernelpolymatrix} from data. This is formalized in the following corollary (see also \cite{Markovsky22}).
\begin{cor}\textup{\cite[Cor. 2]{Alsalti2023_md}}\label{cor_kernelRd}
	Given $w\in\mathscr{B}|_T$ where $\mathscr{B}\in\partial\mathscr{L}_{m,n,\ell}^{q}$, suppose $\mathrm{rank}(\mathscr{H}_d(w))=md+n$ for $d\geq\ell+1$. Then, the coefficients of $R(\sigma)$ in \eqref{eqn_kernelpolymatrix} are given by $R_d\in\mathbb{R}^{pd-n\times qd}$ where $R_d=\mathrm{null}(\mathscr{H}_d(w)^\top)^\top$.
\end{cor}
It was further shown in \cite{Alsalti2023_md} that one can use $R_d$ to obtain a data-based representation (alternative to that in \eqref{GPE_FL}) of the finite-length behavior of the system $\mathscr{B}|_L$. This result is summarized in the following lemma.
\begin{lem}\textup{\cite[Th. 3, Cor. 3]{Alsalti2023_md}}\label{thm_AFL}
	Let the conditions of Corollary~\ref{cor_kernelRd} hold. Then, for any $L\geq d$, $\bar{w}\in\mathscr{B}|_L$ if and only if there exists a vector $\beta\in\mathbb{R}^{mL+n}$ such that
	\vspace{-0.5em}
	\begin{equation}
		P\beta=\bar{w},\label{eqn_efficientFL}
		\vspace{-0.5em}
	\end{equation}
	where $P=\textup{null}(\Gamma)$ and $\Gamma$ is given by
	\vspace{-0.5em}
	\begin{equation}
		\Gamma =\label{Bmatrix}\begin{tikzpicture}[decoration={brace,amplitude=5pt},baseline=(current bounding box.west), scale=0.95, every node/.style={scale=0.9}]
			\node[align=center] at (0,0) {$\begingroup\setlength\arraycolsep{2.5pt}\begin{bmatrix}
					\begin{matrix}
						r_{1,0}\\[-0.75em]
						r_{2,0}\\[-0.75em]
						\vdots\\[-0.75em]
						r_{pd-n,0}
					\end{matrix} & \begin{matrix}
						r_{1,1}\\[-0.75em]
						r_{2,1}\\[-0.75em]
						\vdots\\[-0.75em]
						r_{pd-n,1}
					\end{matrix} & \begin{matrix}
						\cdots\\[-0.75em] \cdots\\[-0.75em] \ddots\\[-0.75em] \cdots
					\end{matrix} & \begin{matrix}
						r_{1,d-1}\\[-0.75em]
						r_{2,d-1}\\[-0.75em]
						\vdots\\[-0.75em]
						r_{pd-n,d-1}
					\end{matrix} &\\[-0.75em]
					& \begin{matrix}
						r_{1,0}\\[-0.75em]
						r_{2,0}\\[-0.75em]
						\vdots\\[-0.75em]
						r_{p,0}
					\end{matrix} & \begin{matrix}
						r_{1,1}\\[-0.75em]
						r_{2,1}\\[-0.75em]
						\vdots\\[-0.75em]
						r_{p,1}
					\end{matrix} & \begin{matrix}
						\cdots\\[-0.75em] \cdots\\[-0.75em] \ddots\\[-0.75em] \cdots
					\end{matrix} & \begin{matrix}
						r_{1,d-1}\\[-0.75em]
						r_{2,d-1}\\[-0.75em]
						\vdots\\[-0.75em]
						r_{p,d-1}
					\end{matrix}\\[-0.75em]
					& &\ddots & \ddots & \ddots & \ddots\\[-0.75em]
					& & & \begin{matrix}
						r_{1,0}\\[-0.75em]
						r_{2,0}\\[-0.75em]
						\vdots\\[-0.75em]
						r_{p,0}
					\end{matrix} & \begin{matrix}
						r_{1,1}\\[-0.75em]
						r_{2,1}\\[-0.75em]
						\vdots\\[-0.75em]
						r_{p,1}
					\end{matrix} & \begin{matrix}
						\cdots\\[-0.75em] \cdots\\[-0.75em] \ddots\\[-0.75em] \cdots
					\end{matrix} & \begin{matrix}
						r_{1,d-1}\\[-0.75em]
						r_{2,d-1}\\[-0.75em]
						\vdots\\[-0.75em]
						r_{p,d-1}
					\end{matrix}
				\,\,\end{bmatrix}\endgroup$};
			\draw[decorate] (1.75,1.1) -- (3.65,-1) node[above=5pt,midway,sloped] {$L-d$ \textup{times}};
		\end{tikzpicture},
	\end{equation}
	where $r_{i,j}\in\mathbb{R}^{1\times q}$ are the elements of $R_d$ in Corollary~\ref{cor_kernelRd}.
\end{lem}
Lemma~\ref{thm_AFL} and the fundamental lemma \cite[Th. 1]{Willems05} both provide data-based representations of the finite-length behavior of an LTI system. This allows for employing the Hankel matrix or the matrix $P$ to infer trajectories over a prediction horizon $L$ in a predictive control scheme. However, there are two important distinctions: \textbf{(D1)} To satisfy the PE condition on the input, Lemma~\ref{thm_AFL} requires $T\geq (m+1)(\ell+n+1)-1$ data points, whereas the fundamental lemma \cite[Th. 1]{Willems05} requires $T\geq (m+1)(L+n)-1$, which depends on $L$. Even if the minimum $T$ is chosen in both cases, then Lemma~\ref{thm_AFL} will always require $(m+1)(L-\ell-1)$ \textit{fewer} samples, for any $L>\ell+1$. \textbf{(D2)} The dimension of $\beta\in\mathbb{R}^{mL+n}$ in \eqref{eqn_efficientFL} is \textit{independent} of $T$, whereas the dimension of $\alpha\in\mathbb{R}^{T-L+1}$ in \eqref{GPE_FL} increases with $T$. In fact, even if the minimum $T$ was chosen in the fundamental lemma \cite[Th. 1]{Willems05}, then the dimension of $\beta$ would still be smaller than $\alpha$ by $mn$.

\begin{rem}\label{remark_mind}
	Although the results of \cite{Alsalti2023_md} hold for the general case of irregularly measured data, in this paper we consider for simplicity that the offline data is complete. For complete and noise-free data, one can already obtain $R_d$ for $d=\ell+1$ if \eqref{eqn_GPE} is satisfied (cf. Corollary~\ref{cor_kernelRd}). Later in Section~\ref{sec_rob_eDDPC} when dealing with noisy data, $d$ assumes the role of a hyperparameter that can be tuned to enhance the performance of the proposed robust eDDPC scheme.
\end{rem}
\vspace{-0.5em}
\section{Nominal eDDPC scheme}
\vspace{-0.5em}
\label{sec_eDDPC}
We now present the efficient data-driven predictive control scheme (eDDPC) in the nominal noise-free setting (initially proposed in our preliminary conference version~\cite{alsalti2023eddpc_ecc}). The goal of eDDPC is to stabilize a (known) equilibrium point of an unknown LTI system, while satisfying input-output constraints. Such a point is defined in terms of the system's inputs and outputs as follows.
\begin{definition}\textup{\cite{Berberich203}}
	A point $w^s$ is an equilibrium of $\mathscr{B}\in\partial\mathscr{L}_{m,n,\ell}^{q}$ if $w'\in(\mathbb{R}^q)^{n+1}$, with $w_k'=w^s$ for all $k\in\mathbb{Z}_{[0,n]}$, is a trajectory of the system, i.e., $w'\in\mathscr{B}|_{n+1}$. We use $w^s_n$ to denote a column vector containing $n$ instances of $w^s$.
\end{definition}
The eDDPC scheme uses the matrix $P$ (cf. \eqref{eqn_efficientFL}) to make the predictions over the horizon\footnote{The length of the predicted trajectories is extended by $n$ instances to account for the initialization step (cf. \eqref{eqn_eDDPC_ini}).} $L+n$. Recall that Lemma~\ref{thm_AFL} implements a few (algebraic) pre-processing steps on the collected data in order to arrive at the matrix $P$ in \eqref{eqn_efficientFL}. In Algorithm~\ref{alg_preprocessing}, we summarize these steps which can be done offline after the data collection phase. Note that the only requirement on the collected data is that it satisfies the rank condition $\textup{rank}(\mathscr{H}_d(w))=md+n$, which can be enforced by design of input; otherwise, the data can be arbitrary. To implement the proposed eDDPC scheme, the following finite-horizon optimal control problem is solved at each time $t$
\vspace{-1em}
\begin{subequations}
	\begin{align}
		\min_{\beta(t),\bar w(t)}\quad 
		& \sum_{k=0}^{L-1}l(\bar{w}_k(t))\label{eqn_eDDPC_cost}\\
		\textrm{s.t.}\quad 
		& \bar w_{[-n,L-1]}(t) = P\beta(t)\label{eqn_eDDPC_dynamics}\\
		& \bar w_{[-n,-1]}(t) = w^{\textup{on}}_{[t-n,t-1]}\label{eqn_eDDPC_ini}\\
		& \bar w_{[L-n,L-1]}(t) = w^s_n\\
		& \bar w_k(t) \in \mathbb{W},\quad \forall k \in \mathbb{Z}_{[0,L-1]}.%
	\end{align}\label{eqn_eDDPC}%
\end{subequations}%
Here, $\bar{w}(t)\in\mathbb{R}^{q(L+n)}$ refers to the predicted input-output trajectories at time $t$, while online measurements are denoted by $w^{\textup{on}}_t$. The stage cost \eqref{eqn_eDDPC_cost} is a quadratic function that penalizes the deviation from the given set point, i.e., $l(\bar{w}_k(t))=\norm*{\bar w_k(t) - w^s}_W^2$, for some $W\succ0$. Finally, $\mathbb{W}$ denotes the constraint set and is defined as
\vspace{-0.5em}
\begin{equation}
	\mathbb{W}\coloneqq\{w=\begin{bsmallmatrix}
		u\\y
	\end{bsmallmatrix}~|~ u\in\mathbb{U}\subseteq\mathbb{R}^m,\,y\in\mathbb{Y}\subseteq\mathbb{R}^p\},\label{eqn_constraintset}\vspace{-0.5em}
\end{equation}
where $\mathbb{U},\mathbb{Y}$ are input and output constraint sets, respectively, with $w^s\in\mathrm{int}(\mathbb{W})$. Once a solution to \eqref{eqn_eDDPC} is found (denoted $\beta^*(t)$ and $\bar{w}^*(t)$), the first instant of the optimal input $\bar{u}_0^*(t)$ is applied to the system and the process is repeated in a receding horizon fashion (see Algorithm~\ref{alg_eDDPC}). Notice that, since $\mathrm{im}(P)=\mathrm{im}(\mathscr{H}_{L+n}(w))$, it follows that eDDPC \eqref{eqn_eDDPC} and existing DDPC schemes that rely on the use of Hankel matrices are equivalent and the resulting closed-loop trajectories of the corresponding schemes are identical (see \cite{alsalti2023eddpc_ecc} for details). As a result, it follows that the proposed eDDPC scheme retains the same theoretical guarantees as the ones shown in \cite{Berberich203} for the nominal case.
\begin{algorithm}[!t]
	\hrule\vspace{-0.5em}
	\caption{Offline data pre-processing for eDDPC}\label{alg_preprocessing}
	\hrule
	\textbf{Input:} Measurements $w\in\mathscr{B}|_T$, where $\mathscr{B}\in\partial\mathscr{L}_{m,n,\ell}^{q}$, satisfying rank$(\mathscr{H}_{d}(w))=md+n$ for $d\geq\ell+1$.
	\begin{itemize}
		\item[1)] Compute $R_d = \mathrm{null}(\mathscr{H}_d(w)^\top)^\top$.
		\item[2)] Use $R_d$ to build $\Gamma$ as in \eqref{Bmatrix}, with $L+n-d$ shifts.
		\item[3)] Obtain $P=\mathrm{null}(\Gamma)$.
	\end{itemize}
	\textbf{Output:} Matrix $P$ where im$(P)=\mathscr{B}|_{L+n}$.
	\hrule
\end{algorithm}%
\begin{algorithm}[!t]
	\hrule\vspace{-0.5em}
	\caption{Nominal eDDPC scheme}\label{alg_eDDPC}
	\hrule
	\textbf{Input:}  Measurements $w\in\mathscr{B}|_T$, where $\mathscr{B}\in\partial\mathscr{L}_{m,n,\ell}^{q}$, satisfying rank$(\mathscr{H}_{d}(w))=md+n$ for $d\geq\ell+1$.\\
	\textbf{Offline phase:} run Algorithm~\ref{alg_preprocessing} to obtain $P$.\\
	\textbf{Online phase:}
	\begin{itemize}
		\item[1.] At time $t$, use measurements $w^{\textup{on}}_{[t-n,t-1]}$ to solve \eqref{eqn_eDDPC}.
		\item[2.] Apply $u^{\mathrm{on}}_t=\bar{u}^*_0(t)$ to the system.
		\item[3.] Set $t=t+1$ and return to Step 1.
	\end{itemize}
	\hrule
\end{algorithm}%

In Section~\ref{sec_rob_eDDPC}, we propose a robust version of \eqref{eqn_eDDPC} in presence of noisy measurements (both in the offline and online phases). To show recursive feasibility and stability of the resulting closed-loop system, we first provide in the next section a novel result on uncertainty quantification in the behavioral framework.
\section{Uncertainty quantification in the behavioral framework}\label{sec_uncertainty_quantification}
We consider output\footnote{Our analysis is also applicable for the case when noise affects both input and output channels. Here, we focus on output noise for simplicity and since this is the standard setting considered in DDPC literature (see, e.g., \cite{Berberich203,Berberich22_IRP}).} measurements which are affected by uniformly bounded additive noise. In particular, one has access to measurements $\widetilde w$ satisfying
\vspace{-0.5em}
\begin{equation}
	\widetilde w_k = \begin{bsmallmatrix}
		u_k\\ \tilde{y}_k
	\end{bsmallmatrix} = \begin{bsmallmatrix}
		u_k\\ y_k
	\end{bsmallmatrix} + \begin{bsmallmatrix}
	0\\ \varepsilon_k
\end{bsmallmatrix} \eqqcolon w_k + \epsilon_k.\label{eqn_noisyw}
\vspace{-0.5em}
\end{equation}
We make the following standard assumption on the noise sequence (compare \cite{Berberich203,Berberich22_IRP,verheijen2023handbook,berberich2024overview} and references therein).
\begin{assumption}\label{asmp_noisydata}
		There exists a known $\bar\varepsilon>0$ such that $\norm*{\epsilon_k}_{\infty}=\norm*{\varepsilon_k}_{\infty}\leq\bar\varepsilon$ for all $k\geq0$.
\end{assumption}
When applying a persistently exciting input to the system and collecting noisy output data, the corresponding data matrix, in general, has $\mathrm{rank}(\mathscr{H}_d(\widetilde w))\geq md+n$ and, hence, one cannot follow Algorithm~\ref{alg_preprocessing} to get $P$. A possible remedy is to first obtain a low rank approximation $\widehat{\mathscr{H}}$ of $\mathscr{H}_d(\widetilde w)$ such that $\mathrm{rank}(\widehat{\mathscr{H}}) = md+n$. Low rank approximation can be done using, e.g., truncated singular-value decomposition (TSVD, cf. \cite{Eckart36}), or using structured low-rank approximation of Hankel matrices (SLRA, cf. \cite{Markovsky08}). To illustrate the use of TSVD, let the singular value decomposition of $\mathscr{H}_{d}(\widetilde w)$ be given by
\vspace{-0.6em}
\begin{equation}
	{\mathscr{H}}_{d}(\widetilde w) = \big[\begin{matrix}
		U & W
	\end{matrix}\big]\textup{diag}(S_1,S_2)\big[\begin{matrix}
	V & Q
\end{matrix}\big]^\top,
\vspace{-0.6em}
\end{equation}
where $S_1 = \mathrm{diag}(s_1({\mathscr{H}}_{d}(\widetilde w)),\ldots,s_{md+n}({\mathscr{H}}_{d}(\widetilde w)))$, $S_2$ contains the remaining singular values and $U,W,V,Q$ are semi-orthonormal matrices of appropriate dimensions. An approximate matrix $\widehat{\mathscr{H}}$ can now be obtained as
\vspace{-0.6em}
\begin{equation}
	\widehat{\mathscr{H}} = US_1V^\top.\label{eqn_TSVD_noisyw}%
	\vspace{-0.7em}
\end{equation}
Since $\mathrm{rank}(\widehat{\mathscr{H}})=md+n$ (by construction), we can now follow the steps in Algorithm~\ref{alg_preprocessing} to obtain a matrix $\widehat{P}$ whose image defines an approximation of the finite-length behavior of the system, i.e., $\mathrm{im}(\widehat{P})\eqqcolon \widehat{\mathscr{B}}|_{L+n}$. In Section~\ref{sec_rob_eDDPC_scheme}, we will use $\widehat{P}$ as a predictor in a robust eDDPC scheme. To later prove stability of such a scheme, we first need to quantify the discrepancy between the (unknown) true behavior $\mathscr{B}|_{L+n}=\mathrm{im}(P)$ and the (known) approximate behavior $\widehat{\mathscr{B}}|_{L+n}=\mathrm{im}(\widehat{P})$.

Discrepancy between behaviors (subspaces) is studied in terms of the \textit{distance} between them \cite{sasane2003}, which is a function of the principal angles (see \cite[Th. 7]{Fazzi23}). Recent works \cite{Padoan22, Fazzi23} define new metrics between \textit{known} behaviors along with methods to compute them based on kernel or data-based representations. However, these results cannot be readily used in our setting, since the true behavior is unknown and, therefore, a bound on the distance must be derived instead. To this end, we provide a novel result on uncertainty quantification in the behavioral framework. In particular, we provide bounds on the angles between $\mathscr{B}|_{L+n}$ and $\widehat{\mathscr{B}}|_{L+n}$ in terms of the noise level~$\bar{\varepsilon}$. This is done by exploiting SVD perturbations (see Section~\ref{sec_SVDperturbations}) and is summarized in the following theorem, which is the main result of this section.
\begin{thm}\label{thm_uncertaintyquantification}
	Given noisy measurements $\widetilde w$ of a trajectory $w\in\mathscr{B}|_T$ as in \eqref{eqn_noisyw} where $\mathscr{B}\in\partial\mathscr{L}_{m,n,\ell}^{q}$, let Assumption~\ref{asmp_noisydata} hold and that $\mathrm{rank}(\mathscr{H}_d(\widetilde w))\geq \textup{rank}(\mathscr{H}_d(w))= md+n$. For any $L\geq d\geq\ell+1$, let $\widehat{\mathscr{B}}|_{L+n}\coloneqq\textup{im}(\widehat{P})$ where $\widehat{P}$ is obtained by Algorithm~\ref{alg_preprocessing} following a TSVD approximation of $\mathscr{H}_{d}(\widetilde w)$. Then,
	\vspace{-0.5em}
	\begin{equation}
		\norm{\sin(\Theta(\widehat{\mathscr{B}}|_{L+n},\mathscr{B}|_{L+n}))}_F\leq \frac{C_{\theta}}{\delta_1\delta_2}\bar{\varepsilon},\label{eqn_uncertaintyquantification}
		\vspace{-0.5em}
	\end{equation}
	where $C_{\theta}=8 \sqrt{qd(L+n-d+1)(pd-n)(T-d+1)}$, $\delta_1=s_{p(L+n)-n}(\widehat\Gamma)$, $\delta_2=s_{md+n}(\widehat{\mathscr{H}})$ and $\sin(\cdot)$ is applied element-wise.
\end{thm}\vspace{-0.5cm}
\begin{pf}
	See Appendix~\ref{appendix_UncertaintyQuantification}.
\end{pf}\vspace{-0.5cm}

Notice that larger values of $\delta_1,\delta_2$ lead to a tighter error bound \eqref{eqn_uncertaintyquantification}. Here, $\delta_2=s_{md+n}(\widehat{\mathscr{H}})$ which, due to TSVD, is also equal to $s_{md+n}(\mathscr{H}_d(\widetilde{w}))>0$. It was shown in \cite[Th. 6]{Coulson23} that a lower bound on $s_{md+n}(\mathscr{H}_d(\widetilde{w}))$ can be guaranteed by suitable design of the input. Simply put, inputs with larger \textit{quantitative levels of PE} result in large values of $\delta_2$, thus reducing the bound in \eqref{eqn_uncertaintyquantification}.

It can further be shown that \eqref{eqn_uncertaintyquantification} goes to zero as $\bar{\varepsilon}\to0$. To see this, notice that as a consequence of Lemma~\ref{thm_Mirsky} applied with $E\coloneqq \widehat{\Gamma} - \Gamma$, $\delta_1=s_{p(L+n)-n}(\widehat{\Gamma})$ can be bounded as $s_{p(L+n)-n}(\Gamma)-\norm{E}_F \leq \delta_1$. Thus, one can write
\vspace{-0.5em}
\begin{align}
	&\begin{aligned}
		s_{p(L+n)-n}(\Gamma) \leq \delta_1 + \norm{E}_F &\stackrel{\eqref{eqn_E_Gamma}}{\leq} \delta_1 + c_1\norm{\widehat{R}_d-R_d}_F\\
		&\stackrel{\eqref{eqn_RdHatRd}}{\leq} \delta_1+ \frac{\rho_1}{\delta_2} \bar\varepsilon
	\end{aligned}\notag\\
	&\implies\,\frac{1}{\delta_1} \leq \frac{\delta_2}{\delta_2 s_{p(L+n)-n}(\Gamma) - \rho_1\bar\varepsilon},\label{eqn_bnd_delta1}
	\vspace{-0.5em}
\end{align}
where $\rho_1 \coloneqq 4c_1\sqrt{2qd(pd-n)(T-d+1)}$. Plugging this back into \eqref{eqn_uncertaintyquantification} results in
\vspace{-0.5em}
\begin{equation}
	\hspace{-0.25mm}\norm{\sin(\Theta(\widehat{\mathscr{B}}|_{L+n},\mathscr{B}|_{L+n}))}_F \leq \frac{C_{\theta}/s_{p(L+n)-n}(\Gamma)}{\delta_2 - \bar\rho_1\bar\varepsilon}\bar{\varepsilon}\label{eqn_uncertaintyquantification_bnd_delta1},
	\vspace{-0.5em}
\end{equation}
where $\bar{\rho}_1 = \rho_1/s_{p(L+n)-n}(\Gamma)$. Using \eqref{eqn_TSVDbound2} and following similar steps that led to \eqref{eqn_bnd_delta1}, we obtain
\vspace{-0.5em}
\begin{equation}
	\frac{1}{\delta_2 - \bar\rho_1\bar\varepsilon}\leq \frac{1}{s_{md+n}(\mathscr{H}_d(w))- (\bar\rho_1+\rho_2) \bar{\varepsilon}}, \label{eqn_bnd_delta2}
	\vspace{-0.5em}
\end{equation}
for $\rho_2\coloneqq 2 \sqrt{qd(T-d+1)}$. Finally, we can further bound \eqref{eqn_uncertaintyquantification_bnd_delta1} using \eqref{eqn_bnd_delta2} as follows
\vspace{-0.5em}
\begin{gather}
	\hspace{-20mm}\norm{\sin(\Theta(\widehat{\mathscr{B}}|_{L+n},\mathscr{B}|_{L+n}))}_F\label{eqn_bnd_angleB_goesto0}\\ \hspace{10mm}\leq\frac{C_{\theta}/s_{p(L+n)-n}(\Gamma)}{s_{md+n}(\mathscr{H}_d(w))- (\bar\rho_1+\rho_2)\bar\varepsilon}\bar{\varepsilon}.\notag
	\vspace{-0.5em}
\end{gather}
From here, it is easy to see that the bound goes to zero as $\bar\varepsilon\to0$, provided that $s_{md+n}(\mathscr{H}_d(w))>(\bar\rho_1+\rho_2)\bar\varepsilon$, where $s_{md+n}(\mathscr{H}_d(w))$ is the smallest non-zero singular value of the noise-free Hankel matrix. Such a condition can be guaranteed by design of input (cf. \cite[Th. 4]{Coulson23}). By combining the results of Lemma~\ref{lemma_newbasis} and Theorem~\ref{thm_uncertaintyquantification} together with \eqref{eqn_bnd_angleB_goesto0}, we obtain the following corollary.
\begin{cor}\label{cor_hatP_P}
	Let the assumptions of Theorem~\ref{thm_uncertaintyquantification} hold and suppose that $s_{md+n}(\mathscr{H}_d(w))>(\bar\rho_1+\rho_2)\bar\varepsilon$. Then, there exists $P$ such that $\mathrm{im}( P)=\mathscr{B}|_{L+n}$ and
	\vspace{-0.5em}
	\begin{equation}
		\label{eqn_hatP_P} 
		\norm{\widehat{P}- P}_F \leq \frac{2\sqrt{m(L+n)+n}C_{\theta}/s_{p(L+n)-n}(\Gamma)}{s_{md+n}(\mathscr{H}_d(w))- (\bar\rho_1+\rho_2)\bar\varepsilon}\bar{\varepsilon},
		\vspace{-0.5em}
	\end{equation}
	where $C_{\theta},\delta_1,\delta_2$ are as in \eqref{eqn_uncertaintyquantification} and $\bar{\rho}_1,{\rho}_2$ are as in \eqref{eqn_bnd_angleB_goesto0}.
\end{cor}
\begin{rem}\label{rem_robustSVDDDPC}
	Similar bounds to \eqref{eqn_uncertaintyquantification} and \eqref{eqn_hatP_P} can also be derived if one directly works with the Hankel matrices of data. This, however, requires the availability of longer sequences of data which is not the setting considered in this paper. Such bounds can be used for stability analysis of, e.g., the robust DDPC scheme from \cite{zhang2022}, but is potentially of interest in applications beyond DDPC.
\end{rem}
\section{Robust eDDPC}\label{sec_rob_eDDPC}
\vspace{-0.5em}
	In this section, we present a robust eDDPC scheme to counteract the effect of noisy data in both online and offline phases
	\vspace{-0.5em}
	\begin{equation}
		\begin{aligned}
			\widetilde{w}_t &= w_t + \epsilon_t, &&t\in\mathbb{Z}_{[0,T-1]},\\
			\widetilde{w}^{\mathrm{on}}_t &= w^{\mathrm{on}}_t + \epsilon^{\mathrm{on}}_t, \quad &&t\in\mathbb{Z}_{\geq 0},
		\end{aligned}\label{eqn_noisyonlinedata}
		\vspace{-0.5em}
	\end{equation}
	where the online noise is also assumed to satisfy Assumption~\ref{asmp_noisydata}, i.e., $\norm*{\epsilon^{\mathrm{on}}_t}_{\infty}\leq\bar\varepsilon$ for all $t\geq0$. Later in Section~\ref{sec_effcom}, we compare its sample and computational requirements against existing robust DDPC schemes from the literature. Finally, in Section~\ref{sec_rob_eDDPC_stability} we prove recursive feasibility and practical exponential stability of the resulting closed-loop system. To that end, we use the results of Corollary~\ref{cor_hatP_P} and follow a suitably modified version of the proof technique presented in \cite{Berberich22_IRP}, which lays a framework to analyze robustness of DDPC schemes based on inherent robustness of nominal MPC schemes \cite[Sec. 3.5]{Rawlings17}.
\subsection{Robust scheme}\label{sec_rob_eDDPC_scheme}
\vspace{-0.5em}
We propose a robust formulation of the eDDPC scheme which uses $\widehat{P}$ from Section~\ref{sec_uncertainty_quantification} as a predictor. Specifically, at time $t$ we use the most recent online noisy measurements $\widetilde w^{\textup{on}}_{[t-n,t-1]}$ to solve the following problem
\vspace{-0.5em}
\begin{subequations}
	\begin{align}
		\min_{\hat{\beta}(t),\hat{w}(t),\hat{\sigma}(t)}\, 
		& \sum_{k=0}^{L-1}l(\widehat{w}_k(t))+ \lambda_{\beta}\bar{\varepsilon}^{\mu_{\beta}}\norm*{\widehat{\beta}(t)}_2^2+ \frac{\lambda_{\sigma}}{\bar{\varepsilon}^{\mu_{\sigma}}}\norm*{\widehat{\sigma}(t)}_2^2\label{eqn_robust_eDDPC_cost}\\
		\mathrm{s.t.}\,\, 
		& \widehat w_{[-n,L-1]}(t) + \widehat{\sigma}(t)= \widehat{P}\widehat{\beta}(t)\label{eqn_robust_eDDPC_dynamics}\\
		& \widehat w_{[-n,-1]}(t) = \widetilde{w}^{\textup{on}}_{[t-n,t-1]}\label{eqn_robust_eDDPC_ini}\\
		& \widehat w_{[L-n,L-1]}(t) = w_n^s\label{eqn_robust_eDDPC_terminal}\\ % or 0
		& \widehat w_k(t) \in \mathbb{W},\quad \forall k \in \mathbb{Z}_{[0,L-1]}.%
	\end{align}\label{eqn_robust_eDDPC}%
\end{subequations}%
To mitigate the effect of noise (both in the online and offline phases), we include a slack variable $\widehat{\sigma}(t)\in\mathbb{R}^{q(L+n)}$ and regularize it in the cost function along with the regressor vector $\widehat{\beta}(t)$ using regularization parameters $\lambda_\beta,\lambda_\sigma,\mu_{\beta},\mu_{\sigma}>0$. To later show recursive feasibility and stability of this scheme, we require that the last two parameters are chosen such that $\mu_{\beta}+\mu_{\sigma}<2$. Such regularization techniques are standard in existing works on robust DDPC (cf., \cite{Berberich22_IRP, Doerfler23, Breschi23c} and many others \cite{verheijen2023handbook,berberich2024overview}). Here, we use $\widehat{w}(t),\widehat{\beta}(t),\widehat{\sigma}(t)$ to denote the decision variables of the robust scheme, to distinguish it from $\bar{w}(t),{\beta}(t)$ that were used for the nominal scheme \eqref{eqn_eDDPC}. The optimal solutions of \eqref{eqn_robust_eDDPC} at time $t$ are denoted by $\widehat{w}^*(t),\widehat{\beta}^*(t),\widehat{\sigma}^*(t)$. Unlike in the nominal case (Section~\ref{sec_eDDPC}), we consider the following assumption on the constraints set (see \eqref{eqn_constraintset}).
\vspace{-0.5em}
\begin{assumption}\label{asmp_polytopic}
	The set $\mathbb{U}$ is a convex compact polytope and $\mathbb{Y}=\mathbb{R}^p$.
\end{assumption}
\vspace{-0.5em}
Methods to address output constraint satisfaction in robust DDPC can be found in, e.g., \cite{Berberich2020,kloppelt22}, which we expect to be applicable to our proposed eDDPC scheme.

\begin{table*}[!t]
	\caption{Analytic comparison between eDDPC and existing DDPC schemes.\\Note that for sDDPC each segment is at least $T_{\textup{seg}}\geq\ell$, and for eDDPC $d$ satisfies $d\geq\ell+1$.}
	\centering
	\begin{tabular}{ | c | c | c | c | c | }
		\hline 	
		& DDPC \cite{Berberich203} & sDDPC \cite{Dwyer23} & SVD-DDPC \cite{zhang2022} & eDDPC\\ \hline
		$T\geq$ & $(m+1)(L+2n)-1$ & $(m+1)(2T_{\textup{seg}}+n)-1$ & $(m+1)(L+2n)-1$ & $(m+1)(d+n)-1$\\ \hline
		$\begin{matrix}
			\textup{dim(regressor)}
		\end{matrix}$ & $T-L-n+1$ & $n_{\textup{seg}}(T-2T_{\textup{seg}}+1)$ & $m(L+n)+n$ & $m(L+n)+n$\\
		\hline
	\end{tabular}
	\label{table_analytic_comparison}
\end{table*}
We consider an $n$ step receding horizon scheme. In particular, once a solution for \eqref{eqn_robust_eDDPC} is found, we apply the first $n$ instances of the optimal input $\widehat{u}^*(t)$. Afterwards, the horizon is shifted by $n$ steps and the procedure is repeated, see Algorithm~\ref{alg_rob_eDDPC}. The reason for considering a multi-step scheme is due to the stability analysis shown later in Section~\ref{sec_rob_eDDPC_stability}, compare \cite{Berberich22_IRP}. Specifically, we show the inherent robustness of the $n$ step scheme with respect to small input disturbances, which is later exploited to show stability of the closed-loop system as in Theorem~\ref{thm_guarantees}. In the following subsection, we discuss the sample and computational efficiency of our proposed robust scheme.
\begin{algorithm}[!t]
	\hrule\vspace{-0.5em}
	\caption{Robust eDDPC scheme}\label{alg_rob_eDDPC}
	\hrule
	\textbf{Input:}  Measurements $\widetilde{w}$ of $w\in\mathscr{B}|_T$, where $\mathscr{B}\in\partial\mathscr{L}_{m,n,\ell}^{q}$, satisfying rank$(\mathscr{H}_{d}(\widetilde{w}))\geq md+n$ for $d\geq\ell+1$.\\
	\textbf{Offline phase:} Obtain a low-rank approximation $\widehat{\mathscr{H}}$ of $\mathscr{H}_d(\widetilde{w})$ such that $\mathrm{rank}(\widehat{\mathscr{H}})=md+n$, and follow Algorithm~\ref{alg_preprocessing} to obtain $\widehat{P}$.\\
	\textbf{Online phase:}
	\begin{itemize}
		\item[1.] At time $t$, use measurements $\widetilde{w}^{\textup{on}}_{[t-n,t-1]}$ to solve \eqref{eqn_robust_eDDPC}.
		\item[2.] Apply $u^{\mathrm{on}}_{[t,t+n-1]}=\widehat{u}^*_{[0,n-1]}(t)$ to the system.
		\item[3.] Set $t=t+n$ and return to Step 1.
	\end{itemize}
	\hrule
\end{algorithm}%

\vspace{-0.5em}
\subsection{Sample and computational efficiency}\label{sec_effcom}
\vspace{-0.5em}
We analytically compare the sample and computational requirements of our proposed robust eDDPC scheme \eqref{eqn_robust_eDDPC} to (i) the DDPC scheme of \cite{Berberich203}, (ii) the segmented DDPC (sDDPC) scheme of \cite{Dwyer23} and (iii) the SVD-based DDPC scheme (SVD-DDPC) of \cite{zhang2022}. Here, sample requirements refer to the minimum number of offline data points required for successful application of each scheme, whereas computational requirements are expressed in terms of the total number of decision variables.

We first discuss sample requirements of all schemes. Recall that the robust eDDPC scheme \eqref{eqn_robust_eDDPC} uses $\widehat{P}$ as a predictor, which can be obtained from short and noisy data. In particular, given $\widetilde{w}$ satisfying $\mathrm{rank}(\mathscr{H}_d(\widetilde{w}))=md+n$ for $d\geq\ell+1$, one can perform a low-rank approximation of it, then follow the steps in Algorithm~\ref{alg_preprocessing} to arrive at $\widehat{P}$. For small noise levels, applying a PE input of order $d+n$ generally results in $\mathrm{rank}(\mathscr{H}_d(\widetilde{w}))\geq md+n$. A necessary condition for this is that the data is of length $T\geq (m+1)(d+n)-1$, which is crucially independent of the prediction horizon length $L$. In contrast, both DDPC and SVD-DDPC schemes \cite{Berberich203,zhang2022} require at least $T\geq(m+1)(L+2n)-1$ samples (corresponding to PE of the input of order $L+n$). For sDDPC \cite{Dwyer23}, the length of each segment must satisfy $T_{\textup{seg}}\geq\ell$ and thus requires $T\geq(m+1)(2T_{\textup{seg}}+n)-1$ data points (see also \cite{alsalti2023eddpc_ecc} for more details). Assuming that the prediction horizon is an integer multiple of $T_{\textup{seg}}$ (i.e., $L=n_{\textup{seg}}T_{\textup{seg}}$ where $n_{\textup{seg}}$ is the number of segments), sDDPC and eDDPC schemes would use the same number of data points only when $T_{\textup{seg}}=\ell=1$. Otherwise, eDDPC always uses less data points. Finally, we emphasize that, unlike other schemes, the robust eDDPC scheme \eqref{eqn_robust_eDDPC} can be applied when the offline data contains irregularly measured samples, provided that $\widehat{R}_d$ can be computed from the available samples (see \cite[Alg. 3]{Alsalti2023_md}). Table~\ref{table_analytic_comparison} summarizes this comparison and highlights the fact that robust eDDPC is the most sample-efficient DDPC scheme.

Regarding computational requirements, we consider the number of decision variables as a key metric and an indicator for the complexity of the optimal control problem. Since predicted inputs and outputs are of the same length for all schemes, the only decision variables that differ in size are the regressor vectors and slack variables. Table~\ref{table_analytic_comparison} lists the dimension of the regressor vector in the optimal control problem of each scheme. The regressor vector of the proposed eDDPC scheme is the smallest (and is equal to that of SVD-DDPC), whereas the size of the regressor vectors of sDDPC and DDPC grows as more offline data is used, i.e., larger $T$. In fact, even if the minimum $T$ is chosen, the regressor vector for eDDPC would still be smaller. Notice that in the nominal setting, eDDPC is the most sample and computationally efficient scheme. This is further highlighted in our preliminary conference version \cite{alsalti2023eddpc_ecc} with a simulation case study on random systems of varying dimensions (system order, number of inputs and outputs).

In the robust setting, slack variables are used to serve different purposes and, hence, each scheme uses a different number of them. For instance, in order to show stability, \cite{Berberich203} uses $p(L+n)$ slack variables to account for noise in the output measurements. Together with the size of the regressor vector listed in Table~\ref{table_analytic_comparison}, this amounts to a total number of $T+(p-1)(L+n)+1$ decision variables which scales with $T$ (when the data is noisy, $T$ is typically chosen larger than the minimum number of points needed). In contrast, the robust eDDPC scheme in \eqref{eqn_robust_eDDPC} uses $q(L+n)$ slack variables (where $q=m+p$) even when considering only output noise. The increased number of the slack variables is due to the pre-processing steps performed when building the matrix $\widehat{P}$. Specifically, when low-rank approximation is performed on the noisy data matrix $\mathscr{H}_d(\widetilde{w})$ (cf. \eqref{eqn_TSVD_noisyw}), all the entries of the resulting approximate matrix $\widehat{\mathscr{H}}$ (and, subsequently $\widehat{P}$) are potentially affected by the output measurement noise. As a result, $q(L+n)$ slack variables are needed when later showing stability and recursive feasibility of the robust eDDPC scheme (see Section~\ref{sec_rob_eDDPC_stability}). Together with the dimension of $\widehat{\beta}(t)$ in \eqref{eqn_robust_eDDPC}, this amounts to a total number of $(q+m)(L+n)+n$ decision variables which does \emph{not} scale with the number of data points. Notice that, eDDPC always uses less decision variables than DDPC once $T\geq T_{\mathrm{min}}+mL$ where $T_{\mathrm{min}}\coloneqq(m+1)(L+2n)-1$ is the minimum number of data points required for the DDPC scheme. The SVD-DDPC scheme \cite{zhang2022} only considers slack variables to enhance closed-loop performance but no stability or robustness guarantees are provided in case of noisy data. In fact, showing closed-loop guarantees for a robust SVD-DDPC scheme (which can be done following similar procedure as in Section~\ref{sec_rob_eDDPC_stability}, compare Remark~\ref{rem_robustSVDDDPC}) would require $q(L+n)$ slack variables for the same reason explained above for the case of eDDPC and, hence, a total number of decision variables equal to eDDPC, i.e., $(q+m)(L+n)+n$. Finally, the sDDPC \cite{Dwyer23} employs $2pL$ slack variables to improve performance but does not provide any stability or robustness guarantees. This, together with the dimension of the regressor vectors employed there, amounts to $n_{\textup{seg}}(T-2T_{\textup{seg}}+1)+2pL$ which also scales with $T$.

Based on the above discussion, we expect that eDDPC and SVD-DDPC achieve similar computation times, followed by DDPC and sDDPC. This trend is more pronounced as $T$ increases. This analysis is further supported empirically by numerical simulations in Section~\ref{sec_examples}.
\vspace{-0.5cm}
\subsection{Stability guarantees}\label{sec_rob_eDDPC_stability}
\vspace{-0.5em}
In this subsection, we show that the robust eDDPC scheme \eqref{eqn_robust_eDDPC} is recursively feasible and that the resulting closed-loop system is practically stable.
\vspace{-0.5em}
\begin{rem}\label{rem_SLRAguarantees}
	To analyze stability of the robust eDDPC scheme \eqref{eqn_robust_eDDPC}, we consider the case in which the predictor $\widehat{P}$ is obtained following a TSVD approximation method and then running Algorithm~\ref{alg_preprocessing}. This enables us to use the results of Corollary~\ref{cor_hatP_P} for the stability analysis. Alternatively, one can use SLRA \cite{Markovsky08} when constructing $\widehat{P}$. In this case, obtaining an analogous result to Corollary~\ref{cor_hatP_P} becomes more complicated since SLRA involves solving a nonlinear optimization problem. Nevertheless, we expect that similar stability and recursive feasibility guarantees hold for an SLRA-based eDDPC scheme. This is supported by our (extensive) simulations in Section~\ref{sec_examples} where such an SLRA-based eDDPC did not encounter feasibility or instability issues. 
\end{rem}
\vspace{-0.5em}
Without loss of generality, we consider here the problem of stabilizing the origin, i.e., $w^s = 0$, as non-zero set points can be considered with slight modifications of the following proofs. Similar to \cite{Berberich22_IRP}, we make the following constraint qualification assumption, which corresponds to requiring that the rows of the equality and active inequality constraints in \eqref{eqn_robust_eDDPC} to be linearly independent.
\begin{assumption}\label{asmp_LICQ}
	Problem \eqref{eqn_eDDPC} satisfies the linear independence constraint qualification (LICQ).
\end{assumption}
Since only input-output data is available, we define the non-minimal state $\xi_t^{\mathrm{on}}\coloneqq\Pi w_{[t-n,t-1]}^{\mathrm{on}}=[\begin{matrix}
	(u_{[t-n,t-1]}^\mathrm{on})^\top & (y_{[t-n,t-1]}^\mathrm{on})^\top
\end{matrix}]^\top$ for some suitable permutation matrix $\Pi$ (similarly, $\widetilde{\xi}_t^{\mathrm{on}}\coloneqq \Pi \widetilde{w}_{[t-n,t-1]}^{\mathrm{on}}$ denotes the noisy extended state, cf. \eqref{eqn_noisyw}). The state-space representation corresponding to $\xi$ is detectable \cite{goodwin2014}. Hence, there exists an input-output-to-state stability (IOSS) Lyapunov function $V_{\mathrm{IOSS}}(\xi^{\mathrm{on}}) = \norm{\xi^{\mathrm{on}}}_X^2$ satisfying
\vspace{-0.5em}
\begin{equation}
	\label{eqn_ioss}
	V_{\mathrm{IOSS}}(\xi_{t+1}^{\mathrm{on}}) - V_{\mathrm{IOSS}}(\xi_t^{\mathrm{on}}) \leq -\frac{1}{2} \norm{\xi_t^{\mathrm{on}}}^2 + c\norm{w_t^{\mathrm{on}}}^2,
	\vspace{-0.5em}
\end{equation}
for some $X\succ0$ and $c > 0$ \cite{cai2008}. Using $V_{\mathrm{IOSS}}$, we define
\vspace{-0.5em}
\begin{equation}
	\label{eqn_LyapFcn}
	V(\xi_t^{\mathrm{on}}) = J_L^*(\xi_t^{\mathrm{on}}) + \frac{1}{c}V_{\mathrm{IOSS}}(\xi_t^{\mathrm{on}}),
	\vspace{-0.5em}
\end{equation}
where ${J}^*_L(\xi_t^{\mathrm{on}})$ denotes the optimal cost of \eqref{eqn_eDDPC}. Notice that, due to Assumption~\ref{asmp_LICQ}, $J_L^*$ is quadratically upper bounded, i.e., there exists $c_{J}>0$ such that $J_L^*(\xi_t^{\mathrm{on}})\leq c_J \norm{\xi_t^{\mathrm{on}}}_2^2$ for any $\xi_t$ for which \eqref{eqn_eDDPC} is feasible, compare \cite{Bemporad02}. The function $V(\xi_t^{\mathrm{on}})$ in \eqref{eqn_LyapFcn} can now be used as a practical Lyapunov function \cite[Def. 2.3]{grune2014} for the robust eDDPC scheme \eqref{eqn_robust_eDDPC} in the presence of noise.

As mentioned earlier, we will proceed to prove recursive feasibility and practical stability following the framework laid in \cite{Berberich22_IRP}. Specifically, the effect of output measurement noise is first translated into an input disturbance of a nominal scheme. Then, by exploiting the inherent robustness of the nominal scheme, one can eventually show practical stability of the eDDPC scheme in \eqref{eqn_robust_eDDPC}. This is summarized in the following theorem.

\begin{thm}\label{thm_guarantees}
	Given noisy measurements $\widetilde w$ of a trajectory $w\in\mathscr{B}|_T$ as in \eqref{eqn_noisyw} satisfying Assumption~\ref{asmp_noisydata}, where $\mathscr{B}\in\partial\mathscr{L}_{m,n,\ell}^{q}$ is controllable, let $u\in(\mathbb{R}^m)^T$ be PE of order $d+n$ for $d\geq\ell+1$. Further, let Assumptions~\ref{asmp_polytopic}-\ref{asmp_LICQ} and the assumptions of Corollary~\ref{cor_hatP_P} hold, and let $\mathscr{B}$ be controlled by Algorithm~\ref{alg_rob_eDDPC} for $L\geq \max\{2n, d\}$. Then, for any $\overline{V}>0$, there exist $\varepsilon^*,c_{V_1},c_{V_2}>0$, $c_{V_3}\in(0,1)$ and $\phi\in\mathcal{K}_{\infty}$ such that, for all initial conditions $\xi_0^{\mathrm{on}}$ satisfying $V(\xi_0^{\mathrm{on}})\leq \overline{V}$ and all $\bar{\varepsilon}\leq\varepsilon^*$, Problem \eqref{eqn_robust_eDDPC} is feasible for $t=in$, $i\in\mathbb{N}$, and the closed-loop system satisfies
	\vspace{-0.5em}
	\begin{align}
			c_{V_1}\norm*{\xi_t^{\mathrm{on}}}_2^2&\leq V(\xi_t^{\mathrm{on}}) \leq c_{V_2}\norm*{\xi_t^{\mathrm{on}}}_2^2,\label{eqn_LyapBnds}\\
			V(\xi_{t+n}^{\mathrm{on}}) &\leq c_{V_3}V(\xi_t^{\mathrm{on}}) + \phi(\bar\varepsilon).\label{eqn_LyapDecay}
	\end{align}
\end{thm}
\vspace{-2em}
\begin{pf}
	See Appendix~\ref{appendix_guarantees}. The proof follows similar steps as the proofs of \cite[Th. IV.1, Prop. IV.1 and Cor. IV.1]{Berberich22_IRP} and consists of the following main steps: \textbf{(i)} Translating the measurement noise to an input disturbance for a nominal DDPC scheme. \textbf{(ii)} Showing that the multi-step nominal scheme is inherently robust with respect to input-disturbances and, \textbf{(iii)} establishing practical stability of the robust DDPC scheme \eqref{eqn_robust_eDDPC}. The important difference of our setting compared to \cite{Berberich22_IRP} is the use of $\widehat{P}$ as a predictor in \eqref{eqn_robust_eDDPC} instead of a Hankel matrix. Due to the use of low-rank approximation when constructing $\widehat{P}$, the effect of measurement noise can now no longer be easily translated into an input disturbance for a nominal scheme as in \cite{Berberich22_IRP}. To this end, in Appendix~\ref{appendix_guarantees1} we employ slack variables for the input as well as the output in order to show the claim. This requires several modifications that make use of the uncertainty quantification results from Theorem~\ref{thm_uncertaintyquantification} and Corollary~\ref{cor_hatP_P}. Once this is established, recursive feasibility and stability follow using similar arguments as in \cite{Berberich22_IRP}, see Appendices \ref{appendix_guarantees2} and \ref{sec_recfeasproof}.
\end{pf}
\vspace{-1em}
Theorem~\ref{thm_guarantees} shows that if the $n$ step robust eDDPC scheme is initially feasible, then it is recursively feasible and the closed-loop system is practically exponentially stable. Specifically, the closed-loop trajectories converge to a region $\mathbb{V}$ around the origin whose size is proportional to the noise level. As in \cite{Berberich22_IRP}, this result is qualitative and quantifying $\mathbb{V}$ without model knowledge can be difficult. The results of Theorem~\ref{thm_guarantees} can potentially be more conservative than the analogous ones presented in \cite{Berberich22_IRP} for a robust DDPC scheme that uses Hankel matrices of noisy data. This is because our analysis uses the uncertainty quantification result in Theorem~\ref{thm_uncertaintyquantification}, for which the corresponding bounds tend to be conservative. In contrast, the uncertainty due to additive noise in \cite{Berberich22_IRP} can easily be separated and upper bounded (due to the use of Hankel matrices). However, as explained below Theorem~\ref{thm_uncertaintyquantification}, larger quantitative orders of PE lead to tighter error bounds in \eqref{eqn_hatP_P}. An important advantage of the robust eDDPC over the robust scheme from \cite{Berberich22_IRP} is that it achieves comparable performance with very little number of data points, for which the scheme from \cite{Berberich22_IRP} cannot even be implemented, as shown in the following section.
\vspace{-0.5em}
\section{Simulations}\label{sec_examples}
\vspace{-0.5em}
In this section, we consider the problem of regulating a (known) non-zero set point of the (linearized) four tank system. We shall compare the performance of the eDDPC scheme \eqref{eqn_robust_eDDPC} against the DDPC and SVD-DDPC schemes from \cite{Berberich22_IRP,zhang2022}, respectively, for different numbers of available noisy offline data points. In the case of eDDPC, we also compare the use of two different low-rank approximation methods in Algorithm~\ref{alg_rob_eDDPC}. When TSVD is used, we refer to the scheme as eDDPC, and when SLRA is used we refer to it as SLRA-eDDPC.

The linearized state space model of the four tank system is given by
\vspace{-0.5em}
\begin{align}
	x_{k+1} &= \begin{bsmallmatrix}
		0.921 & 0 & 0.041 & 0\\
		0 & 0.918 & 0 & 0.033\\
		0 & 0 & 0.924 & 0\\
		0 & 0 & 0 & 0.937
	\end{bsmallmatrix}x_k + \begin{bsmallmatrix}
		0.017 & 0.001 \\ 
		0.001 & 0.023 \\
		0 & 0.061\\
		0.072 & 0
	\end{bsmallmatrix} u_k\notag\\
	y_k &= \begin{bsmallmatrix}
		1 & 0& 0& 0\\
		0 & 1& 0& 0\\
	\end{bsmallmatrix}x_k + \varepsilon_k,\label{eq:fourtank_system}
\end{align}
where $x_k\in\mathbb{R}^4$, $u_k,y_k,\varepsilon_k\in\mathbb{R}^2$ are the state, input, output and measurement noise at time step $k$, respectively. The objective is to use the above DDPC schemes with a prediction horizon $L = 16$ to stabilize the set point $w^s = [(u^s)^\top \, (y^s)^\top]^\top$ where $(u^s)^\top = [1 \,\,\, 1]$ and $(y^s)^\top =[0.65 \,\,\, 0.77]$. Such a setting has been previously considered in the context of DDPC, see \cite{Berberich203,verheijen2023handbook}. Although our theoretical guarantees were obtained for the case where $\mathbb{Y}=\mathbb{R}^p$ (see Assumption~\ref{asmp_polytopic}), here in the simulation example we show that our method is also applicable for the case when both input and output constraints need to be satisfied and are given by $\mathbb{U} = \mathbb{Y} = [-2,2]^2$. Furthermore, we use a quadratic stage cost $l(\bar{w}_k(t))=\norm*{\bar w_k(t) - w^s}_W^2$ with $W = \mathrm{diag}(10^{-2}I_2,\, 3I_2)$.

To collect the offline data, we perform an open-loop simulation by applying a random input and introducing noise sampled from uniform random distributions: $u_k \sim U(-4,4)^2$ and $\varepsilon_k \sim U(-4\cdot10^{-3},4\cdot10^{-3})^2$, respectively. Several such simulations are performed in order to collect data of different lengths\footnote{The values $T\in\{23,35,47,59,71\}$ correspond to the minimum number of data points required to build $\mathscr{H}_d(\widetilde{w})$ for $d\in\{n,2n,\ldots,5n\}$. For DDPC and SVD-DDPC, only lengths $T\in\{71,100,200,300\}$ were considered as the minimum number of data points is $T\geq(m+1)(L+2n)-1=71$.} $T\in\{23,35,47,59,71,100,200,300\}$. This is because, unlike DDPC and SVD-DDPC, our proposed eDDPC is still applicable in scenarios where limited offline data is available. When sufficiently long PE data is available to allow for application of DDPC and SVD-DDPC schemes (specifically $T\geq(m+1)(L+2n)-1$), we also use the complete data to test various cases of the eDDPC which arise from the ability to vary the depth $d$ in Algorithm~\ref{alg_rob_eDDPC}.

Once the data is collected, we can set up the optimization problems for each scheme. Notice that the regularization terms in \eqref{eqn_robust_eDDPC} have the same structure as in the DDPC scheme \cite{Berberich22_IRP}, whereas SVD-DDPC originally included different regularization terms, and the slack variables were only used for the initial conditions (cf. \cite{zhang2022}). To ensure stability and recursive feasibility, we implement SVD-DDPC using the same structure of the cost function as eDDPC, terminal constraints, and use the same number of slack variables as well, i.e., both for the predicted inputs and outputs (compare the discussion in Section~\ref{sec_effcom}). For all schemes, the parameters $\mu_\beta$ and $\mu_\sigma$ were set to $\mu_\beta = \mu_\sigma = 0.5$, thus satisfying the requirement that $\mu_\beta+\mu_\sigma<2$. The value of the regularization parameters $\lambda_\beta,\lambda_\sigma$, however, were varied from one scheme to another and needed to be tuned individually. For the purpose of this example, we empirically determined a pair of ``optimal'' choice of parameters $\lambda_\beta$ and $\lambda_\sigma$ for each scheme by varying the regularization parameters over a grid defined by $\lambda_i \in \{0,10^{-6}, 10^{-5}, \ldots ,10^{4}\}$ for $i=\beta,\sigma$, across all schemes for all different data lengths $T$. For each scheme and each combination $(\lambda_\beta,\lambda_\sigma,T)$, we performed 100 simulations, using different initial conditions and different offline data sets. An optimal pair $(\lambda_\beta^*,\lambda_\sigma^*)$ for each combination is chosen as the one corresponding to the least average accumulated cost defined~as
\vspace{-0.5em}
\begin{equation}
	\label{eq:accumulated_cost}
	\mathcal{J} = \sum\nolimits_{t = 0}^{T_{\textup{sim}}-1} \norm{u_t-u^s}_R^2 + \norm{y_t-y^s}_Q^2,
	\vspace{-0.5em}
\end{equation}
which sums up the weighted closed-loop input-output trajectories over the simulation time $T_{\textup{sim}}=300$. All simulations were carried out using MATLAB 2023b on an Intel Core i7-10700K CPU with 3.80GHz and 64 GB RAM. The optimization problems were solved using the \texttt{quadprog} in Matlab.

\begin{table}[!t]
	\caption{Optimal values for the regularization parameters $\lambda_{\beta}$ and $\lambda_{\sigma}$ for the SLRA-eDDPC scheme for different values of $T$.\label{tab_opt_regul_params}}
	\begin{tabular}{|c | c c c c c c c c |}
		\hline
		$d$ & $4$ & $8$ & $12$ & $16$ & $20$ & $20$ & $20$ & $20$ \\
		\hline
		$T$ & 23 & 35 & 47 & 59 & 71 & 100 & 200& 300\\
		\hline
		$\lambda_{\beta}^*$ & $0$ & $0$ & $0$ & $0.1$ & $0.1$ & $0.01$ & $0.01$ & $0.01$ \\
		\hline
		$\lambda_{\sigma}^*$ & $10^{-4}$ & $10^{-2}$ & $0.1$ & $10$ & $10$ &  $10$ & $10$ &  $10$ \\
		\hline
	\end{tabular}
\end{table}

For space reasons, we only report the resulting optimal regularization parameters for SLRA-eDDPC in Table \ref{tab_opt_regul_params}. It can be observed that $\lambda_{\sigma}^*$ increases with a larger number of data points. A larger regularization parameter corresponds to less relaxation of \eqref{eqn_robust_eDDPC_dynamics}, potentially due to a more accurate predictor which is obtained after performing the necessary pre-processing steps of Algorithm~\ref{alg_preprocessing} on a deeper Hankel matrix, i.e., larger $d$. This suggests that using more offline data yields a more accurate predictor. For $\lambda_{\beta}^*$, it can be seen that this parameter attains small values regardless of the number of data points $T$. A sensitivity analysis of this hyperparameter and its effect on the performance of the eDDPC is illustrated in Figure~\ref{fig_sensitivity_lambdaB}. The results indicate that for values of $\lambda_\beta\leq10^{-3}$, the performance remains largely unchanged across different data lengths, suggesting that the algorithm is relatively insensitive to this hyperparameter. A possible explanation for this (i.e., for not largely regularizing $\widehat\beta(t)$) is that the pre-processing steps used in Algorithm~\ref{alg_preprocessing} have a de-noising effect on the data matrix (similarly also for SVD-DDPC). In contrast, DDPC directly uses the Hankel matrix of noisy data and, hence, regularization of the regressor vector is needed.
\begin{figure}[!t]
	\centering
	\resizebox{0.88\linewidth}{!}{% This file was created by matlab2tikz.
\definecolor{mycolor1}{rgb}{0.00000,0.44700,0.74100}%
\definecolor{mycolor2}{rgb}{0.85000,0.32500,0.09800}%
\definecolor{mycolor3}{rgb}{0.92900,0.69400,0.12500}%
\definecolor{mycolor4}{rgb}{0.49400,0.18400,0.55600}%
\definecolor{mycolor5}{rgb}{0.46600,0.67400,0.18800}%
\definecolor{mycolor6}{rgb}{0.30100,0.74500,0.93300}%
\definecolor{mycolor7}{rgb}{0.63500,0.07800,0.18400}%
\definecolor{mycolor8}{rgb}{0.75000,0.00000,0.75000}%
\begin{tikzpicture}

\begin{axis}[%
	font=\large,
width=4.521in,
height=3.566in,
at={(0.758in,0.481in)},
scale only axis,
xmin=-7.28833697891227,
xmax=0.677663573233667,
xlabel style={font=\large\color{white!15!black}},
xlabel={$\lambda_\beta$},
ymin=35.7376603286046,
ymax=199.478958597183,
ylabel style={font=\large\color{white!15!black}},
ylabel={Accumulated Cost $\mathcal J$},
axis background/.style={fill=white},
xmajorgrids,
ymajorgrids,
legend style={at={(0.03,0.97)}, anchor=north west, legend cell align=left, align=left, draw=white!15!black},
xticklabels={$0$,$0$,$10^{-6}$,$10^{-5}$,$10^{-4}$,$10^{-3}$,$10^{-2}$,$10^{-1}$,$1$,$10$},
]
\addplot [color=mycolor1, line width=1.5pt, mark size=1.8pt, mark=square*, mark options={solid, fill=mycolor1, draw=mycolor1}]
  table[row sep=crcr]{%
-7	70.4254782937665\\
-6	70.4260208986273\\
-5	70.430904821441\\
-4	70.4797900255416\\
-3	70.9719055700178\\
-2	75.5617692842584\\
-1	89.7479403600377\\
0	95.6380275782519\\
1	96.4013016421353\\
};
\addlegendentry{$T=23$}

\addplot [color=mycolor2, line width=1.5pt, mark size=1.8pt, mark=square*, mark options={solid, fill=mycolor2, draw=mycolor2}]
  table[row sep=crcr]{%
-7	70.3257229032483\\
-6	70.3263035420166\\
-5	70.3315314130389\\
-4	70.3840197353297\\
-3	70.929395432987\\
-2	78.028514589077\\
-1	172.099280852451\\
0	215.853088424041\\
};
\addlegendentry{$T=35$}

\addplot [color=mycolor3, line width=1.5pt, mark size=1.8pt, mark=square*, mark options={solid, fill=mycolor3, draw=mycolor3}]
  table[row sep=crcr]{%
-7	61.021180426067\\
-6	61.0214343182248\\
-5	61.0237204270267\\
-4	61.0467301995713\\
-3	61.2862314310194\\
-2	64.5874546501042\\
-1	140.504744092468\\
0	215.853088424041\\
};
\addlegendentry{$T=47$}

\addplot [color=mycolor4, line width=1.5pt, mark size=1.8pt, mark=square*, mark options={solid, fill=mycolor4, draw=mycolor4}]
  table[row sep=crcr]{%
-7	54.1012850480574\\
-6	54.1012638116224\\
-5	54.1010727062272\\
-4	54.0991636315278\\
-3	54.0802848531124\\
-2	53.9098478838851\\
-1	53.1156413308221\\
0	65.8849126391404\\
1	215.853088424041\\
};
\addlegendentry{$T=59$}

\addplot [color=mycolor5, line width=1.5pt, mark size=1.8pt, mark=square*, mark options={solid, fill=mycolor5, draw=mycolor5}]
  table[row sep=crcr]{%
-7	50.6417105681186\\
-6	50.6416964071925\\
-5	50.6415689709703\\
-4	50.6402959256029\\
-3	50.6277233679924\\
-2	50.5194877451346\\
-1	50.1827736603249\\
0	63.8912017045551\\
1	215.853088424041\\
};
\addlegendentry{$T=71$}

\addplot [color=mycolor6, line width=1.5pt, mark size=1.8pt, mark=square*, mark options={solid, fill=mycolor6, draw=mycolor6}]
  table[row sep=crcr]{%
-7	49.191384184062\\
-6	49.1913832608868\\
-5	49.1913749541707\\
-4	49.1912920272678\\
-3	49.1904824941964\\
-2	49.1846276791207\\
-1	49.3151794978156\\
0	63.4718630871486\\
1	215.853088424041\\
};
\addlegendentry{$T=100$}

\addplot [color=mycolor7, line width=1.5pt, mark size=1.8pt, mark=square*, mark options={solid, fill=mycolor7, draw=mycolor7}]
  table[row sep=crcr]{%
-7	49.0599987786715\\
-6	49.0599983405512\\
-5	49.0599943993226\\
-4	49.0599582035689\\
-3	49.0595834820893\\
-2	49.0578316254965\\
-1	49.2196595281432\\
0	63.5447859289082\\
1	215.853088424041\\
};
\addlegendentry{$T=200$}

\addplot [color=mycolor8, line width=1.5pt, mark size=1.8pt, mark=square*, mark options={solid, fill=mycolor8, draw=mycolor8}]
  table[row sep=crcr]{%
-7	49.0575094426194\\
-6	49.0575090119853\\
-5	49.0575051382832\\
-4	49.0574665808504\\
-3	49.0571017560026\\
-2	49.0554035172102\\
-1	49.217732336077\\
0	63.5337111938871\\
1	215.853088424041\\
};
\addlegendentry{$T=300$}

\addplot[only marks, mark=triangle*, mark options={}, mark size=2.1651pt, draw=black, fill=black] table[row sep=crcr]{%
x	y\\
-7	70.4254782937665\\
-7	70.3257229032483\\
-7	61.0211804260669\\
-1	53.1156413308221\\
-1	50.1827736603249\\
-2	49.1846276791207\\
-2	49.0578316254965\\
-2	49.0554035172102\\
};
\addlegendentry{$\lambda_\beta^*$}

\end{axis}

\begin{axis}[%
width=5.833in,
height=4.375in,
at={(0in,0in)},
scale only axis,
xmin=0,
xmax=1,
ymin=0,
ymax=1,
axis line style={draw=none},
ticks=none,
axis x line*=bottom,
axis y line*=left
]
\end{axis}
\end{tikzpicture}%
}
	\caption{Accumulated closed loop costs, averaged over 100 simulations, with varying values of $\lambda_\beta$ and a fixed $\lambda_\sigma=\lambda_\sigma^*$ for the SLRA-eDDPC scheme, across different numbers of data points $T$.}
	\label{fig_sensitivity_lambdaB}
\end{figure}
\begin{rem}
	We include 0 in the hyperparameter space which turned out to be an optimal choice for some depths $d$ (see Table~\ref{tab_opt_regul_params}). Theorem~\ref{thm_guarantees}, however, requires that $\lambda_\beta,\lambda_{\sigma}>0$. To maintain theoretical guarantees, one may choose $\lambda_{\beta}$ to be a small number rather than zero in those cases, which does not largely affect the closed-loop cost (compare Figure~\ref{fig_sensitivity_lambdaB}).
\end{rem}
\begin{figure}[!t]
	\centering
	\resizebox{0.76\linewidth}{!}{% This file was created by matlab2tikz.
\definecolor{mycolor1}{rgb}{0.00000,0.44700,0.74100}%
\definecolor{mycolor2}{rgb}{0.85000,0.32500,0.09800}%
\definecolor{mycolor3}{rgb}{0.92900,0.69400,0.12500}%
\definecolor{mycolor4}{rgb}{0.49400,0.18400,0.55600}%

\begin{tikzpicture}
	
	\begin{axis}[
		font = \large,%
		width=4.521in,
		height=3.566in,
		at={(0.758in,0.481in)},
		scale only axis,
		xmin=0,
		xmax=320,
		xlabel style={font=\color{white!15!black}},
		xlabel={Number of data points $T$},
		ymin=42,
		ytick = {42,46,50,54,58,62,66},
		ymax=66,
		ylabel style={font=\color{white!15!black}},
		ylabel={Accumulated cost $\mathcal J$},
		axis background/.style={fill=white},
		xmajorgrids,
		ymajorgrids,
		legend style={legend cell align=left, align=left, draw=white!15!black},
		ylabel near ticks,
		xlabel near ticks
		]
		\addplot [color=mycolor1, line width=1.5pt, mark=square*, mark options={solid, fill=mycolor1, draw=mycolor1}]
		table[row sep=crcr]{%
			71	62.7261186956168\\
			100	44.633825859516\\
			200	43.6472628405782\\
			300	43.6131799189412\\
		};
		\addlegendentry{DDPC}
		
		\addplot [color=mycolor2, line width=1.5pt, mark=square*, mark options={solid, fill=mycolor2, draw=mycolor2}]
		table[row sep=crcr]{%
%			71	62.7239480989832\\
%			100	43.7689488063365\\
%			200	43.4682364587667\\
%			300	43.4604692568705\\
			71	62.461890679459120\\
			100	43.769228852156090\\
			200	43.468457584575580\\
			300	43.460686536070604\\
		};
		\addlegendentry{SVD-DDPC}
		
		\addplot [color=mycolor4, line width=1.5pt, mark=square*, mark options={solid, fill=mycolor4, draw=mycolor4}]
		table[row sep=crcr]{%
			23	63.4944219756093\\
			35	63.3756946345512\\
			47	63.2755455350638\\
			59	62.2515458116007\\
			71	60.6048619069222\\
			100	43.7589718738009\\
			200	43.4663423517292\\
			300	43.4590661510339\\
		};
		\addlegendentry{eDDPC}
		
		\addplot [color=mycolor3, line width=1.5pt, mark=triangle*, mark options={solid, fill=mycolor3, draw=mycolor3}]
		table[row sep=crcr]{%
			23	63.4848541477201\\
			35	63.3435116789502\\
			47	56.126144799984\\
			59	47.7022937796224\\
			71	44.6995365384273\\
			100	43.6244047495722\\
			200	43.4654656166001\\
			300	43.4623331151731\\
		};
		\addlegendentry{SLRA-eDDPC}
		
	\end{axis}
	
\end{tikzpicture}%
}
	\caption{Accumulated closed-loop costs averaged over $100$ simulations, with varying number of data points $T$ and noise level $\bar \varepsilon= 0.004$ for all schemes.}
	\label{fig_cost_vs_N}
\end{figure}
\begin{table*}[!t]
	\caption{Average computation time per iteration (in milliseconds) for all schemes, averaged over 100 simulations for each data length $T$.}
	\centering
	\begin{tabular}{| c | c c c c c c c c |}
		\hline
		$T$ & 23 & 35 & 47 & 59 & 71 & 100 & 200 & 300\\
		\hline
		DDPC & - & - & - & - & 0.6334 & 0.7791 & 1.2738 & 2.1008 \\
		\hline
		SVD-DDPC & - & - & - & - & 0.9711 & 0.9766 & 0.9668 & 0.9630 \\
		\hline
		eDDPC & 0.6882 & 0.6832 & 0.6841 & 0.7043 & 0.6972 & 0.7423 & 0.7312 & 0.7346\\
		\hline
		SLRA-eDDPC & 0.6959 & 0.6791 & 0.7350 & 0.7560 & 0.7410 & 0.7393 & 0.7300 & 0.7287\\
		\hline
	\end{tabular}
	\label{tab_comptime}
\end{table*}

After tuning the hyperparameters, we proceed to investigate the impact of varying the number of data points $T$ and noise levels $\bar \varepsilon$ on the closed-loop performance in a new simulation. To this end, we consider the same choices for the number of data points as before and, for each such $T$, conducted $100$ simulation experiments using newly collected offline noisy data and new initial conditions. The resulting accumulated closed-loop costs, averaged over $100$ experiments, are presented in Figure~\ref{fig_cost_vs_N}. It can be seen that SLRA-eDDPC outperforms the other schemes when small amounts of data are used. Notice that when $T<71$ DDPC and SVD-DDPC cannot be implemented, whereas SLRA-eDDPC with $T=71$ already achieves comparable performance to that of DDPC and SVD-DDPC that use $T\geq100$ data points. When using TSVD, eDDPC achieves good performance with $100$ data points. DDPC and SVD-DDPC show similar performance to eDDPC for $T\geq 71$, which is the minimum data length for these schemes. As $T$ increases, all schemes show comparable performance. The results in Figure~\ref{fig_cost_vs_N} highlight that (i) the SLRA-eDDPC scheme performs well when small number of noisy data is available offline and (ii) the use of SLRA is better suited than TSVD when dealing with noisy data.

Table~\ref{tab_comptime} reports the average computation time per iteration (in milliseconds) for all four schemes, averaged over 100 different simulations for each data length $T$. It can be seen that both variants of eDDPC in general achieve lower average computation time per iteration compared to DDPC and SVD-DDPC (except for $T=71$, where DDPC is slightly better). Notice that SVD-DDPC achieves a slightly larger average computation time despite having the same number of decision variables as eDDPC and SLRA-eDDPC. One possible explanation for this is that the matrix $\widehat{P}$ used in eDDPC and SLRA-eDDPC is well-conditioned\footnote{This is because the columns of $\widehat{P}$ were obtained as basis for the right null space of $\widehat{\Gamma}$, compare Algorithm~\ref{alg_preprocessing}. Implementing this in Matlab, returns (by default) a matrix with condition number equal to 1.}, which is not necessarily the case for the predictor used in SVD-DDPC. Additionally, notice that the computation times for eDDPC, SLRA-eDDPC and SVD-DDPC remain largely unaffected as $T$ increases. This is consistent with our analysis in Section~\ref{sec_effcom}, where the total number of decision variables for eDDPC and SVD-DDPC was independent of $T$. In contrast, the computation time for the DDPC significantly increases with increasing $T$, due to the (linear) increase in the dimension of the regressor as $T$ increases.

Finally, we investigate the effect of various noise levels on the performance of the different  predictive control schemes. Specifically, we are interested in how the accumulated closed-loop cost varies as the noise in the (offline and online) data increases. To this end, we fix the number of data points at $T=200$ and conducted $100$ simulations for each of the following noise levels $\bar \varepsilon \in \{10^{-3}, 4\cdot 10^{-3}, 7 \cdot 10^{-3},10^{-2}\}$. The average accumulated costs over all 100 runs are shown in Figure \ref{fig:cost_vs_v_max}. As expected, the performance of all four schemes gets worse with increasing noise levels, however, SLRA-eDDPC consistently achieves the lowest cost $\mathcal J$, followed by eDDPC, SVD-DDPC and DDPC. This may be attributed to the de-noising effect associated with the low-rank approximation step done when constructing the approximate predictor $\widehat{P}$.
\vspace{-0.5em}
\section{Conclusions}\label{sec_conclusions}
\vspace{-0.5em}
In this paper, we presented a robust and efficient data-driven predictive control scheme for discrete-time linear time-invariant systems. As with other DDPC schemes, no model knowledge is available and, instead, only noisy input-output data are available from an offline experiment. This scheme is more sample efficient (requires less offline data) compared to existing schemes, and is also computationally efficient. This is due to its reliance on an alternative data-based representation of the finite-length behavior of the system which can be obtained from short (and potentially irregularly measured) noisy data. This makes the proposed eDDPC scheme applicable in cases where existing schemes fail due to limited/missing offline data.

\begin{figure}[!t]
	\centering
	\resizebox{0.8\linewidth}{!}{% This file was created by matlab2tikz.
\definecolor{mycolor1}{rgb}{0.00000,0.44700,0.74100}%
\definecolor{mycolor2}{rgb}{0.85000,0.32500,0.09800}%
\definecolor{mycolor3}{rgb}{0.92900,0.69400,0.12500}%
\definecolor{mycolor4}{rgb}{0.49400,0.18400,0.55600}%
\begin{tikzpicture}

\begin{axis}[%
	font=\large,
width=4.521in,
height=3.566in,
at={(0.758in,0.481in)},
scale only axis,
xmin=0.001,
xmax=0.01,
xlabel style={font=\color{white!15!black}},
xlabel={Noise level $\bar \varepsilon$},
ymin=46.4,
ymax=48.4,
ylabel style={font=\color{white!15!black}},
ylabel={Accumulated cost $\mathcal J$},
axis background/.style={fill=white},
xmajorgrids,
ymajorgrids,
legend style={at={(0.03,0.97)}, anchor=north west, legend cell align=left, align=left, draw=white!15!black},
ylabel near ticks,
xlabel near ticks
]
\addplot [color=mycolor1, line width=1.5pt, mark=square*, mark options={solid, fill=mycolor1, draw=mycolor1}]
  table[row sep=crcr]{%
0.001	46.5231039502331\\
0.004	46.8672178911849\\
0.007	47.5708523463086\\
0.01	48.2645537778971\\
};
\addlegendentry{DDPC}

\addplot [color=mycolor2, line width=1.5pt, mark=square*, mark options={solid, fill=mycolor2, draw=mycolor2}]
  table[row sep=crcr]{%
0.001	46.5099297455867\\
0.004	46.661417570517\\
0.007	46.8463897876807\\
0.01	47.070297250563\\
};
\addlegendentry{SVD-DDPC}

\addplot [color=mycolor4, line width=1.5pt, mark=square*, mark options={solid, fill=mycolor4, draw=mycolor4}]
  table[row sep=crcr]{%
0.001	46.5090610042941\\
0.004	46.6588998940014\\
0.007	46.8392188183117\\
0.01	47.0522024365145\\
};
\addlegendentry{eDDPC}

\addplot [color=mycolor3, line width=1.5pt, mark=triangle*, mark options={solid, fill=mycolor3, draw=mycolor3}]
table[row sep=crcr]{%
	0.001	46.5084986744398\\
	0.004	46.6467903943784\\
	0.007	46.7996667972538\\
	0.01	46.9667350037874\\
};
\addlegendentry{SLRA-eDDPC}

\end{axis}
\end{tikzpicture}%
}
	\caption{Accumulated closed-loop costs, averaged over $100$ simulations, with varying noise levels $\bar \varepsilon$ and $T=200$ data points for all schemes.}
	\label{fig:cost_vs_v_max}
\end{figure}

For our proposed robust eDDPC, we proved recursive feasibility and practical stability of the closed-loop system, unlike recent literature on efficient DDPC that lack theoretical guarantees. To do so, we introduced a novel result on uncertainty quantification in the behavioral framework. In particular, we derived a bound on the angle between two subspaces: the unknown finite-length behavior of the system and its known approximation from noisy input-output data. Under certain conditions, this bound goes to zero as the noise level tends to zero.

To illustrate the performance of this scheme compared to others in the literature, we conducted a simulation case study on a linearized model of four tank system. Our results show that, when sufficiently long PE data is available, the scheme performs similarly to existing ones from the literature. When significantly less data is available, however, none of the other existing DDPC schemes can be applied. In contrast, our eDDPC scheme is still applicable and results in comparable performance to the case when long enough data is available, in the sense that the accumulated closed-loop costs are very close to one another.

Several extensions of the proposed eDDPC scheme can be made. For instance, we considered regulation of constant set points but the scheme can be extended to tracking DDPC as in \cite{berberich2020_trackingsetpoints,Berberich204}. Moreover, output constraint satisfaction can be achieved following, e.g., \cite{Berberich2020,kloppelt22}. Finally, applying the proposed scheme to real-world systems is another interesting venue for future work.
\begin{ack}
	This work has received funding from the European Research Council (ERC) under the European Union’s Horizon 2020 research and innovation programme (grant agreement No 948679).
\end{ack}

\bibliographystyle{ieeetr}
\bibliography{refs}

\appendix
\section{Proof of Lemma~\ref{lemma_newbasis}}\label{appendix_newbasis}
\vspace{-0.5em}
Recall from the discussion below Definition~\ref{def_principalangles} that principal vectors defining the bases of two subspaces (here $\mathrm{im}(U_M)$ and $\mathrm{im}(U_{\widehat{M}})$) always exist. Let the columns of $\overline{U}_M$ and $\overline{U}_{\widehat M}$ be such principal vectors satisfying $\mathrm{im}(\overline{U}_M)=\mathrm{im}(U_M)$ and $\mathrm{im}(\overline{U}_{\widehat M})=\mathrm{im}(U_{\widehat{M}})$, respectively. Consequently, there exists a matrix $G=[g_1\,\,\cdots\,\,g_r]\in\mathbb{R}^{r\times r}$ such that $U_{\widehat{M}} = \overline{U}_{\widehat{M}} G$. Recall that the columns of $U_{\widehat{M}}=\big[\begin{matrix}u_{\widehat{M},1}&\cdots&u_{\widehat{M},r}\end{matrix}\big]$ are orthonormal and, as a result, the following holds
\vspace{-0.5em}
\begin{align}
		u_{\widehat{M},i}^\top u_{\widehat{M},j} = 0\, &\implies g_i^\top \overline{U}_{\widehat{M}}^\top \overline{U}_{\widehat{M}} g_j = g_i^\top g_j=0,\quad \forall i\neq j,\notag\\
		u_{\widehat{M},i}^\top u_{\widehat{M},i} = 1\, &\implies g_i^\top \overline{U}_{\widehat{M}}^\top \overline{U}_{\widehat{M}} g_i = g_i^\top g_i=1,\label{eqn_normal_u}
\end{align}
where $\overline{U}_{\widehat{M}}^\top \overline{U}_{\widehat{M}}=I_r$ holds by definition of the principal vectors. Notice that \eqref{eqn_normal_u} also implies that the matrix $G$ is orthogonal. Letting $\widetilde{U}_{M}\coloneqq \overline{U}_{M} G$ we obtain a basis for $\mathrm{im}(U_M)$ as desired. Since $\widetilde{U}_{M}$ is the product of two orthogonal matrices, it is orthogonal as well.

Now, it remains to be shown that \eqref{eqn_newbasis} holds. Using (the given) $U_{\widehat{M}}$ and (the constructed) $\widetilde{U}_{M}$, we write
\vspace{-0.5em}
\begin{align}
		\norm{U_{\widehat M}-\widetilde{U}_{M}}_F &= \norm{ \overline{U}_{\widehat{M}} G - \overline{U}_M G}_F=\norm{(\overline{U}_{\widehat M}-\overline{U}_M) G}_F\notag\\
		&\hspace{-20mm}\leq \norm{\overline{U}_{\widehat M}-\overline{U}_M}_F\norm{G}_F\label{eqn_Gremoved}\leq \sqrt{r}\norm{\overline{U}_{\widehat M}-\overline{U}_M}_F,%\\&\notag
\end{align}
where the last inequality follows from $\norm*{G}_F = \sqrt{\sum\nolimits_{i=1}^{r}\norm*{g_i}_2^2}=\sqrt{r}$ since $g_i^\top g_i=1$ as in \eqref{eqn_normal_u}. Now consider a pair of vectors $\bar u_{M,j},\bar u_{\widehat{M},j}$ which are columns of $\overline{U}_M, \overline{U}_{\widehat{M}}$, respectively. By definition of the 2-norm, we have $\norm{\bar u_{\widehat{M},j} - \bar u_{M,j}}_2^2=(\bar u_{\widehat{M},j} - \bar u_{M,j})^\top(\bar u_{\widehat{M},j} - \bar u_{M,j})$, or
\vspace{-0.5em}
\begin{equation*}
	\norm{\bar u_{\widehat{M},j} - \bar u_{M,j}}_2^2 = 2(1-\bar u_{\widehat{M},j}^\top \bar u_{M,j})= 2(1-\cos\theta_j),
	\vspace{-0.5em}
\end{equation*}
where $\bar u_{M,j}^\top \bar u_{M,j}=\bar u_{\widehat{M},j}^\top \bar u_{\widehat{M},j}=1$ and $\cos\theta_j = \bar u_{\widehat{M},j}^\top \bar u_{M,j}$ hold since the vectors $\bar u_{\widehat{M},j},\bar u_{M,j}$ are principal vectors (cf. \eqref{eqn_PA}). Using the identity $\sin^2\left(\frac{\theta_j}{2}\right)=\frac{1-\cos\theta_j}{2}$ we get
\vspace{-0.5em}
\begin{equation}
	\norm{\bar u_{\widehat{M},j}-\bar u_{M,j}}_2^2 = 4\sin^2(\theta_j/2)\leq 4\sin^2\theta_j,
	\vspace{-0.5em}
\end{equation}
where the last inequality holds for all $\theta_j\in[0,\frac{2\pi}{3}]$, and hence also for $\theta_j\in[0,\frac{\pi}{2}]$ (which are the limits of $\theta_j$ as defined in Definition~\ref{def_principalangles}). Summing over $j\in\{1,\ldots,r\}$ on both sides and taking the square root, we obtain
\vspace{-0.5em}
\begin{equation*}
	\sqrt{\sum\limits_{j=1}^r \norm*{\bar u_{\widehat{M},j}-\bar u_{M,j}}_2^2} \leq \sqrt{\sum\limits_{j=1}^{r}4\sin^2\theta_j}=2 \sqrt{\sum\limits_{j=1}^{r}\sin^2\theta_j}.
	\vspace{-0.5em}
\end{equation*}
Notice that the leftmost side corresponds to the Frobenius norm of the difference between the bases $\overline{U}_{\widehat{M}}$ and $\overline{U}_M$, while the right hand side corresponds to the Frobenius norm of the matrix $\sin(\Theta(\textup{im}(U_M),\textup{im}(U_{\widehat{M}})))$. Hence, $\norm{\overline{U}_{\widehat{M}}-\overline{U}_M}_F \leq 2\norm{\sin (\Theta(\mathrm{im}(U_M),\mathrm{im}(U_{\widehat{M}})))}_F$. Plugging this back into \eqref{eqn_Gremoved} results in \eqref{eqn_newbasis} which completes the proof. \hfill$\blacksquare$
\vspace{-0.5em}
%%%%%%%%%%%%%%%%%%%%%%%%%%%%%%%%%%%%%%%%%%%%%%%%%%%%%%%%%%%%
\section{Proof of Theorem~\ref{thm_uncertaintyquantification}}\label{appendix_UncertaintyQuantification}
\vspace{-0.5em}
Since $\mathrm{rank}(\mathscr{H}_d(\widetilde w))\geq \textup{rank}(\mathscr{H}_d(w))= md+n$, then one can perform a TSVD approximation as in \eqref{eqn_TSVD_noisyw} to obtain $\widehat{\mathscr{H}}$ with $\mathrm{rank}(\widehat{\mathscr{H}})=md+n$. Moreover, a basis for the left null space of the two matrices are given by $R_d=\mathrm{null}(\mathscr{H}_d(w)^\top)^\top$ and $\widehat{R}_d=\mathrm{null}(\widehat{\mathscr{H}}^\top)^\top$, respectively. Using ($L+n-d$) shifts of $R_d$ and $\widehat{R}_d$, one can build $\Gamma$ and $\widehat{\Gamma}$ as in \eqref{Bmatrix}, the SVD of which is denoted by
\vspace{-0.5em}
\begin{equation}
	\begin{aligned}
		\Gamma = \begin{bmatrix}U_{\Gamma} & W_{\Gamma}\end{bmatrix}\textup{diag}(S_{\Gamma},0)\begin{bmatrix}
			V_{\Gamma} & Q_{\Gamma}
		\end{bmatrix}^\top,\\
		\widehat{\Gamma} = \begin{bmatrix}U_{\widehat{\Gamma}} & W_{\widehat{\Gamma}}\end{bmatrix}\textup{diag}(S_{\widehat\Gamma},0)\begin{bmatrix}
			V_{\widehat{\Gamma}} & Q_{\widehat{\Gamma}}
		\end{bmatrix}^\top,
	\end{aligned}
\vspace{-0.5em}
\end{equation}
where $S_{\Gamma}=\mathrm{diag}(s_1(\Gamma),\ldots,s_{p(L+n)-n}(\Gamma))$ and $S_{\widehat{\Gamma}}=\mathrm{diag}(s_1(\widehat{\Gamma}),\ldots,s_{p(L+n)-n}(\widehat{\Gamma}))$ (both of which have rank $p(L+n)-n$ by construction, see also \cite{Alsalti2023_md} for details). Further, $U_{i},W_{i},V_{i},Q_{i}$ (for $i=\{\Gamma,\widehat{\Gamma}\}$) are semi-orthonormal matrices of appropriate dimensions. By Lemma~\ref{thm_wedin},\footnote{As mentioned before Lemma~\ref{thm_wedin}, the bound in \eqref{eqn_Wedin} holds for all four singular subspaces, see \cite{Stewart91}.}
\vspace{-0.5em}
\begin{equation}
	\norm{\sin (\Theta(\mathrm{im}(Q_\Gamma),\mathrm{im}(Q_{\widehat{\Gamma}})))}_F \leq \sqrt{2}/\delta_1\norm*{E}_F,\label{eqn_boundingQgamma}
	\vspace{-0.5em}
\end{equation}
where $\delta_1=s_{p(L+n)-n}(\widehat{\Gamma})>0$ (which holds since $\widehat{\Gamma}$ has rank $p(L+n)-n$ as discussed above) and $E\coloneqq \widehat{\Gamma}-\Gamma$. By properties of the SVD, it holds that $\mathrm{im}(Q_{\Gamma})=\mathrm{ker}(\Gamma)=\mathrm{im}(P)=\mathscr{B}|_{L+n}$ where the last equality holds by Lemma~\ref{thm_AFL} (see \cite[Th. 3]{Alsalti2023_md} for more details). Similarly, it holds that $\mathrm{im}(Q_{\widehat{\Gamma}})=\mathrm{ker}(\widehat{\Gamma})=\mathrm{im}(\widehat P)=\widehat{\mathscr{B}}|_{L+n}$, where the last equality holds by definition of $\widehat{\mathscr{B}}|_{L+n}$ as in the theorem statement. Therefore, we can write \eqref{eqn_boundingQgamma} as
	\vspace{-0.5em}
\begin{equation}
	\norm*{\sin(\Theta(\widehat{\mathscr{B}}|_{L+n},\mathscr{B}|_{L+n}))}_F\leq \sqrt{2}/\delta_1\norm*{E}_F.\label{eqn_WedinAppliedToP}
		\vspace{-0.5em}
\end{equation}
Moreover, $\norm*{E}_F = \norm{\widehat{\Gamma}-\Gamma}_F= \sqrt{\sum\nolimits_{i,j}\norm{e_{i,j}}^2}$ is given by
\begin{align}
	&\norm*{E}_F =\\
	&\sqrt{\scalebox{0.9}{$\sum\limits_{i=1}^{pd-n}\sum\limits_{j=0}^{d-1}$}\norm{\widehat{r}_{i,j}-r_{i,j}}^2 + (L+n-d)\scalebox{0.9}{$\sum\limits_{i=1}^{p}\sum\limits_{j=0}^{d-1}$}\norm{\widehat{r}_{i,j}-r_{i,j}}^2}.\notag
\end{align}
Since $\sum\limits_{i=1}^{p}\sum\limits_{j=0}^{d-1}\norm{\widehat{r}_{i,j}-r_{i,j}}^2\leq \sum\limits_{i=1}^{pd-n}\sum\limits_{j=0}^{d-1}\norm{\widehat{r}_{i,j}-r_{i,j}}^2$, we can write
\begin{align}
	\norm*{E}_F&\leq \sqrt{(L+n-d+1)\scalebox{0.9}{$\sum\limits_{i=1}^{pd-n}\sum\limits_{j=0}^{d-1}$}\norm{\widehat{r}_{i,j}-r_{i,j}}^2}\notag\\
	&= \underbrace{\sqrt{L+n-d+1}}_{\coloneqq c_1}\norm*{\widehat{R}_d-R_d}_F.\label{eqn_E_Gamma}
\end{align}
Plugging back in \eqref{eqn_WedinAppliedToP}, we obtain
\vspace{-0.5em}
\begin{equation}
	\norm{\sin(\Theta(\widehat{\mathscr{B}}|_{L+n},{\mathscr{B}}|_{L+n}))}_F \leq \frac{\sqrt{2} c_1}{\delta_1} \norm*{\widehat{R}_d-R_d}_F.\label{eqn_fullbound1}
	\vspace{-0.5em}
\end{equation}
Now consider the SVD of $\mathscr{H}_d(w)$ and $\widehat{\mathscr{H}}$
	\vspace{-0.5em}
\begin{equation}
	\begin{aligned}
		\mathscr{H}_d(w) &= \begin{bmatrix}
			U_{H} & W_{H}
		\end{bmatrix}\textup{diag}(S_{H},0)\begin{bmatrix}
			V_{H} & Q_{H}
		\end{bmatrix}^\top,\\
		\widehat{\mathscr{H}} &= \begin{bmatrix}
			U_{\widehat H} & W_{\widehat H}
		\end{bmatrix}\textup{diag}(S_{\widehat H},0)\begin{bmatrix}
			V_{\widehat H} & Q_{\widehat H}
		\end{bmatrix}^\top,
	\end{aligned}
	\vspace{-0.5em}
\end{equation}
where $S_{H}=\mathrm{diag}(s_1(\mathscr{H}_d(w)),\ldots,s_{md+n}(\mathscr{H}_d(w)))$, $S_{\widehat{H}}=\mathrm{diag}(s_1(\widehat{\mathscr{H}}),\ldots,s_{md+n}(\widehat{\mathscr{H}}))$, $U_{i},W_{i},V_{i},Q_{i}$ (for $i=\{H,\widehat{H}\}$) are semi-orthonormal matrices of appropriate dimensions and $W_H$ is such that (cf. Lemma~\ref{lemma_newbasis})
\begin{equation}
	\resizebox{\columnwidth}{!}{$\norm*{W_{\widehat H} - W_{H}}_F \leq 2\sqrt{pd-n}\norm*{\sin(\Theta(\textup{im}(W_{\widehat H}),\textup{im}(W_{H})))}_F.$}
	\label{eqn_SVDofHw}
\end{equation}
By Lemma~\ref{thm_wedin}, the following holds
\begin{equation}
	\norm*{\sin(\Theta(\textup{im}(W_{\widehat H}),\textup{im}(W_{H})))}_F \leq \frac{\sqrt{2}}{\delta_2}\norm{\widehat{\mathscr{H}}-\mathscr{H}_d(w)}_F,
	\label{eqn_rd-rdhat}
\end{equation}
where $\delta_2 = s_{md+n}(\widehat{\mathscr{H}})=s_{md+n}({\mathscr{H}}_d(\widetilde w))$, which is strictly positive since $\mathrm{rank}({\mathscr{H}}_d(\widetilde w))\geq\mathrm{rank}({\mathscr{H}}_d(w))=md+n>0$. Notice that the norm on the right hand side of \eqref{eqn_rd-rdhat} can be further bounded by
\begin{align}
		&\norm{\widehat{\mathscr{H}}-\mathscr{H}_d(w)}_F= \norm{\widehat{\mathscr{H}} - \mathscr{H}_d(\widetilde w) + \mathscr{H}_d(\widetilde w) - \mathscr{H}_d(w)}_F\notag\\
		&\leq \norm{\widehat{\mathscr{H}} -\mathscr{H}_d(\widetilde w)}_F + \norm{\mathscr{H}_d(\widetilde w) - \mathscr{H}_d(w)}_F\\
		&= \norm{\widehat{\mathscr{H}} -\mathscr{H}_d(\widetilde w)}_F + \norm{\mathscr{H}_d(\epsilon)}_F.\notag
\end{align}
The first term on the right hand side corresponds to the difference between a matrix and its TSVD approximation, which is known to be bounded by the sum of the truncated singular values \cite{Stewart91}, hence
\begin{align}
	\norm*{\widehat{\mathscr{H}}-\mathscr{H}_d(w)}_F&\leq \sqrt{\scalebox{0.9}{$\sum\nolimits_{i=md+n+1}^{\mathrm{rank}(\mathscr{H}_d(\widetilde w))}$} (s_i(\mathscr{H}_d(\widetilde w)))^2}\notag\\&\quad+ \norm*{\mathscr{H}_d(\epsilon)}_F.\label{eqn_TSVDbound}
\end{align}
Note that by Lemma~\ref{thm_Mirsky}, the following holds $\norm*{\mathscr{H}_d(\epsilon)}_F^2 \geq \sum\nolimits_{i=1}^{\mathrm{rank}(\mathscr{H}_d(\widetilde w))} (s_i(\mathscr{H}_d(\widetilde w)) - s_i(\mathscr{H}_d(w)))^2$, or
\begin{align*}
	\norm*{\mathscr{H}_d(\epsilon)}_F^2 &\leq \scalebox{0.95}{$\sum\nolimits_{i=1}^{md+n}$} (s_i(\mathscr{H}_d(\widetilde w)) - s_i(\mathscr{H}_d(w)))^2 \\
	&+ \scalebox{0.95}{$\sum\nolimits_{i=md+n+1}^{\mathrm{rank}(\mathscr{H}_d(\widetilde w))}$} (s_i(\mathscr{H}_d(\widetilde w)))^2,
\end{align*}
where in the last step we exploited the fact that rank$(\mathscr{H}_d(w))=md+n$ and hence $s_i(\mathscr{H}_d(w))=0$ for $i\geq md+n+1$. It is easy to see from here that $\scalebox{0.95}{$\sum\nolimits_{i=md+n+1}^{\mathrm{rank}(\mathscr{H}_d(\widetilde w))}$} (s_i(\mathscr{H}_d(\widetilde w)))^2 \leq \norm*{\mathscr{H}_d(\epsilon)}_F^2$. Taking the square root and substituting in \eqref{eqn_TSVDbound} yields
\begin{equation}
		\norm{\widehat{\mathscr{H}}-\mathscr{H}_d(w)}_F \leq 2\norm*{\mathscr{H}_d(\epsilon)}_F\leq 2 \sqrt{qd(T-d+1)} \,\bar{\varepsilon},\label{eqn_TSVDbound2}
\end{equation}
for $q=m+p$, where the last inequality holds by definition of the Frobenius norm along with the fact that $\norm*{\epsilon_k}_{\infty}\leq\bar\varepsilon$ for all $k\geq0$ (cf. \eqref{eqn_noisyw}). Finally, we substitute back in \eqref{eqn_rd-rdhat} to obtain
\begin{equation}\label{eqn_Wangle}
	\norm*{\sin(\Theta(\textup{im}(W_{\widehat H}),\textup{im}(W_{H})))}_F \leq \frac{2\sqrt{2qd(T-d+1)}}{\delta_2}\bar{\varepsilon}.
\end{equation}

Recall that $\widehat R_d$ and ${R}_d$ form a basis for the left null spaces of $\widehat{\mathscr{H}}$ and $\mathscr{H}_d(w)$ respectively, and thus $\widehat R_d= W_{\widehat H}^\top $ and $R_d=W_{H}^\top $. This, together with \eqref{eqn_SVDofHw} and the fact that $\norm{W_{\widehat H} - W_{H}}_F = \norm{W_{\widehat H}^\top - W_{H}^\top}_F$, allow us to write
\begin{align}
		\norm*{\widehat{R}_d - R_d}_F &\leq 2\sqrt{pd-n}\norm*{\sin(\Theta(\textup{im}( W_{\widehat{H}}),\textup{im}(W_H)))}_F\notag\\
		&\hspace{-7mm}\stackrel{\eqref{eqn_Wangle}}{\leq} \frac{4\sqrt{2qd(pd-n)(T-d+1)} }{\delta_2} \bar{\varepsilon}.\label{eqn_RdHatRd}
\end{align}
Substituting back in \eqref{eqn_fullbound1} completes the proof.\hfill$\blacksquare$
	\vspace{-0.5em}
%%%%%%%%%%%%%%%%%%%%%%%%%%%%%%%%%%%%%%%%%%%%%%%%%%%%%%%%%%%%
\section{Proof of Theorem~\ref{thm_guarantees}}\label{appendix_guarantees}
	\vspace{-0.5em}
\subsection{Translating online measurement noise to an input disturbances to the nominal scheme}\label{appendix_guarantees1}
	\vspace{-0.5em}
We start by showing that the online measurement noise in the robust scheme \eqref{eqn_robust_eDDPC} can be viewed as an input disturbance to the nominal scheme~\eqref{eqn_eDDPC}. The steps followed here are adaptations from the proof of \cite[Th. IV.1]{Berberich22_IRP}.

Assume that $V(\xi_t^{\mathrm{on}})\leq\overline{V}$ (this will be established recursively later in Section~\ref{sec_recfeasproof}). By definition of $V(\xi_t^{\mathrm{on}})$ (see \eqref{eqn_LyapFcn}), it follows that $J_L^*(\xi_t^{\mathrm{on}}) \leq V(\xi_t^{\mathrm{on}}) \leq \overline{V}$. Now, we proceed by defining a feasible candidate solution for \eqref{eqn_robust_eDDPC} based on the optimal solution of \eqref{eqn_eDDPC}. Specifically, let
\begin{align}
	\widehat{w}(t) &\coloneqq \begin{bmatrix}
		\widetilde{w}^{\textup{on}}_{[t-n,t-1]} \\ \bar{w}_{[0,L-1]}^*(t)
	\end{bmatrix} \stackrel{\eqref{eqn_noisyonlinedata}}{=} \begin{bmatrix}
		w^{\textup{on}}_{[t-n,t-1]} + \epsilon_{[t-n,t-1]}^{\mathrm{on}} \\ \bar{w}_{[0,L-1]}^*(t)
	\end{bmatrix}\notag\\ &\stackrel{\eqref{eqn_eDDPC_ini}}{=} \bar{w}^*_{[-n,L-1]}(t) + \begin{bsmallmatrix}
		\epsilon_{[t-n,t-1]}^{\mathrm{on}}\\ 0_{qL\times1}
	\end{bsmallmatrix}.\label{eqn_candidate1}
\end{align}
Moreover, let $\widehat{\beta}(t)=\beta^*(t)$ where $\beta^*(t)$ satisfies $P\beta^*(t)=\bar{w}^*(t)$ (cf. \eqref{eqn_eDDPC_dynamics}) for some $P$ satisfying \eqref{eqn_hatP_P}. Based on the definitions of $\widehat{w}(t),\widehat{\beta}(t)$, we define the following candidate solution for $\widehat{\sigma}(t)$ as the one which makes \eqref{eqn_robust_eDDPC_dynamics} holds, i.e., $\widehat{\sigma}(t) = \widehat{P}\widehat{\beta}(t) - \widehat{w}(t)$. In particular,
\begin{equation}
	\begin{aligned}
		\widehat{\sigma}(t)&\stackrel{\eqref{eqn_candidate1}}{=} \widehat{P}\widehat{\beta}(t) - \bar{w}^*_{[-n,L-1]}(t)-\begin{bsmallmatrix}
			\epsilon_{[t-n,t-1]}^{\mathrm{on}}\\ 0_{qL\times1}
		\end{bsmallmatrix}\\
		&\stackrel{\eqref{eqn_eDDPC_dynamics}}{=}\widehat{P}\widehat{\beta}(t)-P\beta^*(t)-\begin{bsmallmatrix}
			\epsilon_{[t-n,t-1]}^{\mathrm{on}}\\ 0_{qL\times1}
		\end{bsmallmatrix},
	\end{aligned}
\end{equation}
or $\widehat{\sigma}(t)=(\widehat{P} - P)\widehat{\beta}(t)-\begin{bsmallmatrix}
	\epsilon_{[t-n,t-1]}^{\mathrm{on}}\\ 0_{qL\times1}
\end{bsmallmatrix}$. Taking the norm on both sides allows us to write
\begin{equation}
	\begin{aligned}
		\norm{\widehat{\sigma}(t)}_2^2 &\leq 2\norm{\widehat{P}-P}_2^2\norm{\widehat{\beta}(t)}_2^2 + 2\norm{\epsilon_{[t-n,t-1]}^{\mathrm{on}}}_2^2\\
		&\leq 2\norm{\widehat{P}-P}_F^2\norm{\widehat{\beta}(t)}_2^2 + 2np\bar\varepsilon^2,
	\end{aligned}\label{eqn_boundSigmaHat_pre}
\end{equation}
where the second inequality holds by standard norm equivalences. Recall from Corollary~\ref{cor_hatP_P} that
\begin{equation}
	\norm{\widehat{P}-P}_F \leq \frac{2\sqrt{m(L+n)+n}C_{\theta}/s_{p(L+n)-n}(\Gamma)}{s_{md+n}(\mathscr{H}_d(w))- (\bar\rho_1+\rho_2)\bar\varepsilon}\bar{\varepsilon}.
\label{eqn_hatP_P_recalled}
\end{equation}
For a fixed $\varepsilon^*$ satisfying $0\leq \varepsilon^*<\frac{s_{md+n}(\mathscr{H}_d(w))}{(\bar\rho_1+\rho_2)}$, there exists some (uniform and sufficiently large) constant $c_2>0$ such that for all $\bar\varepsilon\in[0,\varepsilon^*]$, the following holds $\norm{\widehat{P}-P}_F \leq c_2\bar\varepsilon$ (which in turn also implies $\norm{\widehat{P}-P}_F^2 \leq c_2^2\bar\varepsilon^2$). Plugging this back into \eqref{eqn_boundSigmaHat_pre}, we obtain
\begin{equation}
	\begin{aligned}
		\norm{\widehat{\sigma}(t)}_2^2 &\leq 2c_2^2\bar\varepsilon^2 \norm{\widehat{\beta}(t)}_2^2+c_3\bar\varepsilon^2,
	\end{aligned}\label{eqn_boundSigmaHat}
\end{equation}
where we have defined $c_3=2np$ for convenience. 

Now, let $\widehat{J}_L(\widetilde{\xi}_t^{\mathrm{on}})$ denote the cost of \eqref{eqn_robust_eDDPC} associated with the candidate solutions $\widehat{w}(t),\widehat{\beta}(t),\widehat{\sigma}(t)$ above. By optimality, it holds that $\widehat{J}^*_L(\widetilde{\xi}_t^{\mathrm{on}})\leq\widehat{J}_L(\widetilde{\xi}_t^{\mathrm{on}})$ where $\widehat{J}^*_L(\widetilde{\xi}_t^{\mathrm{on}})$ denotes the optimal cost of \eqref{eqn_robust_eDDPC}. Notice that the only difference between $\widehat{J}_L(\widetilde{\xi}_t^{\mathrm{on}})$ and the corresponding optimal cost of \eqref{eqn_eDDPC} (i.e., ${J}^*_L(\xi_t^{\mathrm{on}})$) is the regularization terms of the robust eDDPC Problem \eqref{eqn_robust_eDDPC} (see \eqref{eqn_eDDPC_cost},\eqref{eqn_robust_eDDPC_cost} and \eqref{eqn_candidate1}). Together with $\widehat{J}^*_L(\widetilde{\xi}_t^{\mathrm{on}})\leq\widehat{J}_L(\widetilde{\xi}_t^{\mathrm{on}})$, we write
\begin{align*}
	\hspace{-1mm}\widehat{J}^*_L(\widetilde{\xi}_t^{\mathrm{on}})\hspace{-1mm} - \hspace{-1mm} {J}^*_L(\xi_t^{\mathrm{on}}) &\leq \lambda_{\beta} \bar \varepsilon^{\mu_\beta} \norm{\widehat \beta(t)}^2_2 + \frac{\lambda_{\sigma}}{\bar \varepsilon^{\mu_\sigma}}\norm{\widehat \sigma(t)}^2_2\\
	&\hspace{-8mm}\stackrel{\eqref{eqn_boundSigmaHat}}{\leq} \hspace{-2mm}(\lambda_{\beta} \bar \varepsilon^{\mu_\beta}\hspace{-1mm}+\hspace{-1mm}\lambda_{\sigma}2c_2^2\bar \varepsilon^{2-\mu_\sigma}) \norm{\widehat \beta(t)}^2_2+\hspace{-1mm}\lambda_{\sigma}c_3\bar\varepsilon^{2-\mu_{\sigma}}.
\end{align*}
Recall again that $\widehat{\beta}(t)=\beta^*(t)$. This, together with \eqref{eqn_eDDPC_dynamics} allows us to write $\widehat{\beta}(t)=P^{\dagger}\bar{w}^*(t)$, where $P^\dagger$ is a left inverse of $P$ (which exists since $P$ has full column rank, cf. Lemma~\ref{thm_AFL}). Plugging this expression in the above bound, we obtain $\widehat{J}^*_L(\widetilde{\xi}_t^{\mathrm{on}}) \leq {J}^*_L(\xi_t^{\mathrm{on}}) + (\lambda_{\beta} \bar \varepsilon^{\mu_\beta}+\lambda_{\sigma}2c_2^2\bar \varepsilon^{2-\mu_\sigma}) \norm{ P^{\dagger} }_{2}^{2} \norm{\bar{w}^*(t) }^2_2 +\lambda_{\sigma}c_3\bar\varepsilon^{2-\mu_\sigma}$. Notice that, by \eqref{eqn_eDDPC_ini}, one can write $\norm*{\bar{w}^*(t)}_2^2 = \norm{w^{\mathrm{on}}_{[t-n,t-1]}}_2^2 + \norm{\bar{w}_{[0,L-1]}^*(t)}_2^2$, with the first term being bounded as follows
\[
\begin{aligned}
	\norm{w^{\mathrm{on}}_{[t-n,t-1]}}_2^2 &= \norm{\Pi^{-1}\xi_t^{\mathrm{on}}}_2^2 \\
	&\leq \frac{\norm{\Pi^{-1}}_2^2}{\lambda_{\min}(X)}V_{\mathrm{IOSS}}(\xi_t^{\mathrm{on}}) \leq \frac{c\norm{\Pi^{-1}}_2^2}{\lambda_{\min}(X)}\overline{V},
\end{aligned}
\]
whereas the second term is bounded by $\norm{\bar{w}_{[0,L-1]}^*(t)}_2^2\leq\frac{\norm{\bar{w}^*_{[0,L-1]}(t)}_{\overline{W}_L}^2}{\lambda_{\min}(\overline{W}_L)}=\frac{J_L^*(\xi_t^{\mathrm{on}})}{\lambda_{\min}(\overline{W}_L)}\leq\frac{\overline{V}}{\lambda_{\min}(\overline{W}_L)}$, with $\overline{W}_L\coloneqq\mathrm{diag}(W,\ldots,W)$ concatenated $L$ times. Plugging everything back gives
\begin{align}
	\widehat{J}^*_L(\widetilde{\xi}_t^{\mathrm{on}}) &\leq {J}^*_L(\xi_t^{\mathrm{on}}) + (\lambda_{\beta} \bar \varepsilon^{\mu_\beta}+\lambda_{\sigma}2c_2^2\bar \varepsilon^{2-\mu_\sigma}) \norm{ P^{\dagger} }_{2}^{2} \,\, \times \notag\\
	&\quad \left(\frac{c\norm{\Pi^{-1}}_2^2}{\lambda_{\min}(X)}+\frac{1}{\lambda_{\min}(\overline{W}_L)}\right)\overline{V}+\lambda_{\sigma}c_3\bar\varepsilon^{2-\mu_\sigma}\notag\\
	&\eqqcolon {J}^*_L(\xi_t^{\mathrm{on}}) + \phi_1(\bar \varepsilon),\label{eqn_upperboundcostrob}
\end{align}
where $\phi_1 \in \mathcal{K}_\infty$ due to $\mu_{\beta}+\mu_{\sigma}<2$. This bound will be used in the following subsections to establish an upper bound on $\norm{\widehat u^*(t) - \bar u^*(t)}$.
	\vspace{-0.5em}
\subsubsection{Bound on $\norm{\widehat u^*(t) - \breve{u}^*(t)}$}
	\vspace{-0.5em}
Consider the following auxiliary optimization problem
\begin{subequations}
	\label{eqn_auxiliary_eDDPC}
	\begin{align}
		\min_{\breve\beta(t), \breve w(t)}\quad 
		& \sum_{k=0}^{L-1}\norm{\breve w_k(t)}_{W}^2 + \lambda_{\beta} \bar \varepsilon^{\mu_\beta} \norm{\breve \beta(t)}^2_2 \label{eqn_auxiliary_eDDPC_cost_function}\\
		\textrm{s.t.}\quad 
		& \breve{w}_{[-n,L-1]}(t) + \breve\sigma_1 = P\breve\beta(t) \label{eqn_auxiliary_eDDPC_pred_model}\\
		& \breve w_{[-n,-1]}(t) = w^{\textup{on}}_{[t-n,t-1]} + \breve\sigma_2 \label{eqn_auxiliary_eDDPC_init_constr}\\
		& \breve w_{[L-n,L-1]}(t) = 0_{qn\times 1} \label{eqn_auxiliary_eDDPC_term_constr}\\
		& \breve w_k(t) \in \mathbb{W},\quad \forall k \in \mathbb{Z}_{[0,L-1]},
	\end{align}
\end{subequations}
with the parameter $\breve\sigma$ defined as
\begin{equation}
	\label{eqn_sigma_tilde}
	\breve \sigma = \begin{bmatrix}
		\breve \sigma_1 \\ \breve \sigma_2
	\end{bmatrix} \coloneqq \begin{bmatrix}
		\widehat \sigma^*(t) - (\widehat{P}-P) \widehat \beta^*(t) \\ \epsilon^{\textup{on}}_{[t-n,t-1]}
	\end{bmatrix}.
\end{equation}
It can be easily verified that a candidate solution to \eqref{eqn_auxiliary_eDDPC} is given by the optimal solution of \eqref{eqn_robust_eDDPC} (which exists since \eqref{eqn_robust_eDDPC} is feasible as shown above), i.e., $\breve w(t) = \widehat{w}^*(t)$ and $\breve\beta(t)=\widehat\beta^*(t)$. Denote the corresponding cost associated with this candidate solution to \eqref{eqn_auxiliary_eDDPC} by $\breve J_L(\xi_t^{\mathrm{on}})$ and let the optimal solutions and the corresponding optimal cost be denoted by $\breve w^*(t),\breve\beta^*(t)$ and $\breve J_L^*(\xi_t^{\mathrm{on}})$, respectively. By optimality, it holds that
\begin{align}
	\breve J_L^*(\xi_t^{\mathrm{on}}) &\leq \breve J_L(\xi_t^{\mathrm{on}})\stackrel{\eqref{eqn_auxiliary_eDDPC_cost_function}}{=}\sum\nolimits_{k=0}^{L-1}\norm{\breve w_k(t)}_{W}^2 + \lambda_{\beta} \bar \varepsilon^{\mu_\beta} \norm{\breve \beta(t)}^2_2\notag\\
	&= \sum\nolimits_{k=0}^{L-1}\norm{\widehat w^*_k(t)}_{W}^2 + \lambda_{\beta} \bar \varepsilon^{\mu_\beta} \norm{\widehat \beta^*(t)}^2_2\notag\\
	&\stackrel{\eqref{eqn_robust_eDDPC_cost}}{=} \widehat{J}_L^*(\widetilde{\xi}_t^{\mathrm{on}}) - \frac{\lambda_{\sigma}}{\bar\varepsilon^{\mu_{\sigma}}}\norm*{\widehat{\sigma}^*(t)}_2^2.\label{eqn_aux_candidate_sol}
\end{align}
Similarly, one can define a candidate solution for \eqref{eqn_robust_eDDPC} in terms of the optimal solution of \eqref{eqn_auxiliary_eDDPC} (which exists according to \eqref{eqn_aux_candidate_sol}). In particular, define candidate solutions for \eqref{eqn_robust_eDDPC} as $\widehat{w}(t) = \breve{w}^*(t),\, \widehat{\beta}(t)=\breve{\beta}^*(t)$ and
\begin{equation}
	\begin{aligned}
		\widehat{\sigma}(t) &= (\widehat{P}-P)\widehat{\beta}(t) + \breve{\sigma}_1\\
		&\stackrel{\eqref{eqn_sigma_tilde}}{=} \widehat \sigma^*(t) + (\widehat{P}-P)(\widehat{\beta}(t) - \widehat \beta^*(t)).
	\end{aligned}\label{eqn_sigmacand_sigmaopt}
\end{equation}
Denote the corresponding cost associated with this candidate solution to \eqref{eqn_robust_eDDPC} as $\widehat{J}_L'(\widetilde\xi_t^{\mathrm{on}})$. Notice now that the cost $\widehat{J}_L'(\widetilde\xi_t^{\mathrm{on}})$ differs from the optimal cost of \eqref{eqn_auxiliary_eDDPC} by the regularization term involving the slack variable, i.e.,
\begin{equation}
	\widehat{J}_L'(\widetilde\xi_t^{\mathrm{on}}) - \breve J_L^*(\xi_t^{\mathrm{on}}) = \frac{\lambda_{\sigma}}{\bar\varepsilon^{\mu_\sigma}}\norm*{\widehat{\sigma}(t)}_2^2.\label{eqn_cost_difference_robcand_auxopt}
\end{equation}
Furthermore, since \eqref{eqn_robust_eDDPC} is strongly convex in $\widehat{u}$, there exists $c_4>0$ such that $\norm*{\widehat{u}^*(t) - \widehat{u}(t)}_2^2 \leq c_4(\widehat{J}'_L(\widetilde{\xi}_t^{\mathrm{on}}) - \widehat{J}_L^*(\widetilde{\xi}_t^{\mathrm{on}}))\stackrel{\eqref{eqn_aux_candidate_sol}}{\leq} c_4( \widehat{J}'_L(\widetilde{\xi}_t^{\mathrm{on}}) - \breve{J}_L^*(\xi_t^{\mathrm{on}}) - \frac{\lambda_{\sigma}}{\bar\varepsilon^{\mu_{\sigma}}}\norm*{\widehat{\sigma}^*(t)}_2^2 )$. Using the fact that $\widehat{u}(t)=\breve{u}^*(t)$ (see definition before \eqref{eqn_sigmacand_sigmaopt}), together with \eqref{eqn_cost_difference_robcand_auxopt} this implies that
\begin{equation*}
	\norm*{\widehat{u}^*(t) - \breve{u}^*(t)}_2^2 \leq c_4\frac{\lambda_{\sigma}}{\bar\varepsilon^{\mu_{\sigma}}}( \norm*{\widehat{\sigma}(t)}_2^2 - \norm*{\widehat{\sigma}^*(t)}_2^2 ).
\end{equation*}
Using \eqref{eqn_sigmacand_sigmaopt} and $\norm*{a}_2^2 - \norm*{b}_2^2 \leq \norm*{a-b}_2^2 + 2\norm*{a-b}_2 \norm*{b}_2$ for $a,b\in\mathbb{R}$, we have $\norm{\widehat{u}^*(t) - \breve{u}^*(t)}_2^2 \leq c_4\frac{\lambda_{\sigma}}{\bar\varepsilon^{\mu_{\sigma}}}( \norm{\widehat{\sigma}(t)}_2^2 - \norm{\widehat{\sigma}^*(t)}_2^2 )$, or
\begin{align*}
	\norm{\widehat{u}^*(t) - \breve{u}^*(t)}_2^2&\leq c_4\frac{\lambda_{\sigma}}{\bar\varepsilon^{\mu_{\sigma}}} \left( \norm{\widehat{P}-P}_2^2\norm{\widehat{\beta}(t) - \widehat \beta^*(t)}_2^2 \right.\\ 
	&\hspace{-2mm}\left.+ 2\norm{\widehat{P}-P}_2\norm{\widehat{\beta}(t) - \widehat \beta^*(t)}_2\norm*{\widehat{\sigma}^*(t)}_2 \right).
\end{align*}
Now, we use the result of Corollary~\ref{cor_hatP_P} to bound the term $\norm{\widehat{P}-P}_2\leq\norm{\widehat{P}-P}_F$ which, as in the discussion below \eqref{eqn_hatP_P_recalled}, can be bounded by $c_2\bar\varepsilon$ for any $\bar\varepsilon\in[0,\varepsilon^*]$ for some fixed $\varepsilon^*<\frac{s_{md+n}(\mathscr{H}_d(w))}{(\bar\rho_1+\rho_2)}$. Moreover, notice that $\norm{\widehat{\beta}^*(t)}_2^2\leq\frac{\widehat{J}_L^*(\widetilde{\xi}_t^{\mathrm{on}})}{\lambda_{\beta}\bar\varepsilon^{\mu_{\beta}}}$ and  $\norm{\widehat{\sigma}^*(t)}_2^2\leq\frac{\bar\varepsilon^{\mu_{\sigma}}\widehat{J}_L^*(\widetilde{\xi}_t^{\mathrm{on}})}{\lambda_{\sigma}}$ hold by the optimal cost of \eqref{eqn_robust_eDDPC}. Furthermore, $\widehat{\beta}(t)=\breve{\beta}^*(t)$ (see before \eqref{eqn_sigmacand_sigmaopt}) which implies that $\norm{\widehat{\beta}(t)}_2^2=\norm{\breve{\beta}^*(t)}_2^2\leq \frac{\breve{J}^*_L(\xi_t^{\mathrm{on}})}{\lambda_{\beta}\bar\varepsilon^{\mu_{\beta}}}\stackrel{\eqref{eqn_aux_candidate_sol}}{\leq}\frac{\widehat{J}_L^*(\widetilde{\xi}_t^{\mathrm{on}})}{\lambda_{\beta}\bar\varepsilon^{\mu_{\beta}}}$. Collecting all this together results in
\begin{align}
		&\norm{\widehat{u}^*(t) - \breve{u}^*(t)}_2^2\\ 
		&\leq \frac{4\lambda_{\sigma}c_4c_2^2\bar\varepsilon^2}{\lambda_{\beta}\bar\varepsilon^{\mu_{\sigma}+\mu_{\beta}}}\widehat{J}_L^*(\widetilde{\xi}_t^{\mathrm{on}}) + \frac{ 4\lambda_{\sigma}c_4c_2\bar\varepsilon }{ \sqrt{\lambda_{\beta}\lambda_{\sigma}\bar\varepsilon^{\mu_{\sigma}+\mu_{\beta}}} } \widehat{J}_L^*(\widetilde{\xi}_t^{\mathrm{on}})\notag\\
		&\eqqcolon \left(c_5\bar\varepsilon^{2-\mu_{\beta}-\mu_{\sigma}} + c_6\bar\varepsilon^{0.5(2-\mu_{\sigma}-\mu_{\beta})}\right)\widehat{J}_L^*(\widetilde{\xi}_t^{\mathrm{on}})\notag\\
		&\stackrel{\eqref{eqn_upperboundcostrob}}{\leq} \left(c_5\bar\varepsilon^{2-\mu_{\beta}-\mu_{\sigma}} + c_6\bar\varepsilon^{0.5(2-\mu_{\sigma}-\mu_{\beta})}\right)({J}^*_L(\xi_t^{\mathrm{on}}) + \phi_1(\bar \varepsilon))\notag\\
		&\leq\left(c_5\bar\varepsilon^{2-\mu_{\beta}-\mu_{\sigma}} + c_6\bar\varepsilon^{0.5(2-\mu_{\sigma}-\mu_{\beta})}\right)(\overline{V}+ \phi_1(\bar \varepsilon)).\notag
\end{align}
Finally, taking the square root on both sides results in
\begin{align}
		&\norm*{\widehat{u}^*(t) - \breve{u}^*(t)}_2\leq \phi_2(\bar\varepsilon)\label{eqn_bndoptinputs1}\\
		&\eqqcolon\sqrt{\left(c_5\bar\varepsilon^{2-\mu_{\beta}-\mu_{\sigma}} + c_6\bar\varepsilon^{0.5(2-\mu_{\sigma}-\mu_{\beta})}\right)(\overline{V}+ \phi_1(\bar \varepsilon))}\notag
\end{align}
where $\phi_2\in\mathcal{K}_{\infty}$ due to $\mu_{\beta}+\mu_{\sigma}<2$.
	\vspace{-0.5em}
\subsubsection{Bound on $\norm{\breve{u}^*(t) - u^{\prime*}(t)}$}
	\vspace{-0.5em}
Consider a modification of the minimization problem \eqref{eqn_auxiliary_eDDPC} where $\breve{\sigma}\equiv0$. Such a problem has the same structure as \eqref{eqn_eDDPC} but with one additional term in the cost function, namely $\lambda_{\beta}\bar{\varepsilon}^{\mu_{\beta}}\norm{\breve\beta(t)}_2^2$. Let the optimal solution of such a problem at time $t$ be denoted by $u^{\prime*}(t)$, which exists as implied by the feasibility of \eqref{eqn_eDDPC} at time $t$ (see above). To obtain a bound on $\norm{\breve{u}^*(t) - u^{\prime*}(t)}$, we can follow the same arguments in \cite{Berberich22_IRP}. In particular, there exists $c_7>0$ such that
\begin{align*}
	&\norm{\breve{u}^*(t) - u^{\prime*}(t)}_2 \leq c_7 \norm*{\breve{\sigma}}_2 \leq c_7 \left(\norm*{\breve{\sigma}_1}_2 + \norm*{\breve{\sigma}_2}_2\right)\\
	&\stackrel{\eqref{eqn_sigma_tilde}, \eqref{eqn_aux_candidate_sol}}{\leq} c_7 \sqrt{\frac{\bar\varepsilon^{\mu_{\sigma}}}{\lambda_{\sigma}}\widehat{J}_L^*(\widetilde{\xi}_t^{\mathrm{on}})}+ c_7c_2\bar\varepsilon\scalebox{0.95}{$\sqrt{\frac{\widehat{J}_L^*(\widetilde{\xi}_t^{\mathrm{on}})}{\lambda_{\beta}\bar\varepsilon^{\mu_{\beta}}}}$}+ c_7pn\bar{\varepsilon}.
\end{align*}
This, together with \eqref{eqn_upperboundcostrob} and the fact that ${J}^*_L(\xi_t^{\mathrm{on}})\leq\overline{V}$, results in (for some $c_8,c_9,c_{10}>0$)
	\vspace{-0.5em}
\begin{equation}
	\norm{\breve{u}^*(t) - u^{\prime*}(t)}_2 \leq c_8\bar{\varepsilon} + c_9\bar{\varepsilon}^{\mu_{\sigma}/2} + c_{10}\bar{\varepsilon}^{1-\mu_{\beta}/2}.\label{eqn_bndoptinputs2}
		\vspace{-0.5em}
\end{equation}
\subsubsection{Bound on $\norm{u^{\prime*}(t) - \bar u^*(t)}$}
	\vspace{-0.5em}
Recall again that $u^{\prime*}(t)$ and $\bar u^*(t)$ correspond to the optimal solutions of two different optimization problems which share the same structure, with the only difference being that the former has an additional term in the cost function; namely $\lambda_{\beta}\bar{\varepsilon}^{\mu_{\beta}}\norm{\breve\beta(t)}_2^2$. Following the same arguments as in \cite{Berberich22_IRP}, such a problem can be reformulated into a strongly convex quadratic program (due to Assumption~\ref{asmp_polytopic}) and a bound of the form $\norm{u^{\prime*}(t) - \bar u^*(t)}\leq c_{11}\bar{\varepsilon}^{\mu_{\beta}/2}$, for some $c_{11}>0$, is derived. We omit the steps here for brevity, since the analysis follows similar steps. Combining this together with \eqref{eqn_bndoptinputs1} and \eqref{eqn_bndoptinputs2} yields
\begin{align}
		\hspace{-2mm}\norm{\widehat u^*(t) - \bar u^*(t)}_2 &\leq \norm{\widehat u^*(t) - \breve u^*(t)}_2 + \norm{\breve u^*(t) - u^{\prime*}(t)}_2\notag\\
		&+ \norm{u^{\prime*}(t) - \bar u^*(t)}_2\leq \phi_3(\bar{\varepsilon}),\label{eqn_bndoptinputs}%\notag\\&
\end{align}
for some $\phi_3\in\mathcal{K}_{\infty}$, which results from the sum of the three $\mathcal{K}_{\infty}$ functions (in $\bar{\varepsilon}$).
	\vspace{-0.5em}
\subsection{Recursive feasibility and inherent robustness of the multi-step nominal eDDPC scheme}\label{appendix_guarantees2}
	\vspace{-0.5em}
We now consider that the system $\mathscr{B}$ is controlled by an $n$-step nominal eDDPC scheme (as in \eqref{eqn_eDDPC}), but where the input applied to the system is perturbed from the optimal solution, i.e., $u^{\mathrm{on}}_{[t,t+n-1]} = \bar{u}^*_{[0,n-1]}(t) + \delta_{[t,t+n-1]}$, where $\norm*{\delta_t}_2\leq\bar\delta$ for all $t\geq0$. The idea is that one can view the robust scheme with measurement noise \eqref{eqn_robust_eDDPC} as a nominal scheme \eqref{eqn_eDDPC} with bounded input disturbance as shown in \eqref{eqn_bndoptinputs}. If such a scheme is feasible at time $t$, recursive feasibility can be established using standard arguments from model-based MPC, i.e., by defining a candidate solution at the next time step using the previously optimal solution and suitably appending the predicted inputs by a deadbeat controller (which exists due to controllability assumption and that $0\in\mathrm{int}(\mathbb{W})$). For brevity, we omit this here as it follows similar steps as in \cite[Prop. IV.1]{Berberich22_IRP}. Practical stability is then established using Lyapunov arguments. In particular, the Lyapunov function $V(\xi_t^{\mathrm{on}})$ in \eqref{eqn_LyapFcn} can be bounded as follows
\begin{equation}
	c_{V_1}\norm*{\xi_t^{\mathrm{on}}}_2^2 \leq V(\xi_t^{\mathrm{on}}) \leq 
	c_{V_2}\norm*{\xi_t^{\mathrm{on}}}_2^2,\label{eqn_LyapBnd1}
\end{equation}
where $c_{V_1}\coloneqq \frac{\lambda_{\mathrm{min}}(X)}{c}$ and $c_{V_2}\coloneqq \left(c_{J} + \frac{\lambda_{\mathrm{max}}(X)}{c}\right)$. For the $n$ step decrease condition, notice that
\begin{align}
	&V(\xi_{t+n}^{\mathrm{on}}) - V(\xi_t^{\mathrm{on}})\label{eqn_LyapDecay1}\\
	&=J_L^*(\xi_{t+n}^{\mathrm{on}}) - J_L^*(\xi_{t}^{\mathrm{on}}) + \frac{1}{c}\left(V_{\mathrm{IOSS}}(\xi_{t+n}^{\mathrm{on}}) - V_{\mathrm{IOSS}}(\xi_{t}^{\mathrm{on}})\right)\notag\\
	&\resizebox{\columnwidth}{!}{$\leq -\norm*{w_{[t,t+n-1]}^{\mathrm{on}}}_{\overline{W}_n}^{2} + \phi_4(\bar\delta)+ \frac{1}{c}\left(V_{\mathrm{IOSS}}(\xi_{t+n}^{\mathrm{on}}) - V_{\mathrm{IOSS}}(\xi_{t}^{\mathrm{on}})\right),$}\notag
\end{align}
where $\overline W_{n} = \mathrm{diag}(W,\cdots,W)$. The first term in \eqref{eqn_LyapDecay1} appears due to the definition of the shifted candidate solutions at time $t+n$, whereas the second term (with $\phi_4\in\mathcal{K}_{\infty}$) accounts for the input disturbances. The last term in \eqref{eqn_LyapDecay1} can be bounded by repeatedly applying IOSS arguments (see \eqref{eqn_ioss}) over $n$ steps to obtain
\begin{align}
	&\resizebox{\columnwidth}{!}{$V_{\mathrm{IOSS}}(\xi_{t+n}^{\mathrm{on}}) - V_{\mathrm{IOSS}}(\xi_{t}^{\mathrm{on}})\leq -\norm*{\xi_{[t,t+n-1]}^{\mathrm{on}}}_2^2 + c \norm*{w_{[t,t+n-1]}^{\mathrm{on}}}_{\overline W_n}^2$}\notag\\
	&\stackrel{\eqref{eqn_LyapBnd1}}{\leq} -\frac{1}{c_{V_2}} V(\xi_t^{\mathrm{on}})-\norm*{\xi_{[t+1,t+n-1]}^{\mathrm{on}}}_2^2 + c \norm*{w_{[t,t+n-1]}^{\mathrm{on}}}_{\overline W_n}^2\notag
\end{align}
Dropping the second term (since it is non-positive), and plugging back into \eqref{eqn_LyapDecay1}, we obtain
\begin{equation*}
		V(\xi_{t+n}^{\mathrm{on}})\leq (1-\frac{1}{c c_{V_2}})V(\xi_t^{\mathrm{on}}) + \phi_4(\bar\delta)
\end{equation*}
where $(1-\frac{1}{c c_{V_2}}) < 1$. Clearly, there exists some $c_{V_3}\in(0,1)$ such that $(1-\frac{1}{c c_{V_2}})\leq c_{V_3}$ and, hence,
\begin{equation}
	V(\xi_{t+n}^{\mathrm{on}})\leq c_{V_3}V(\xi_t^{\mathrm{on}}) + \phi_4(\bar\delta).\label{eqn_LyapDecay4}
\end{equation}
	\vspace{-1em}
\subsection{Practical stability of the $n$-step robust eDDPC}\label{sec_recfeasproof}
	\vspace{-0.5em}
As shown above, an $n$-step robust scheme based on \eqref{eqn_robust_eDDPC} can be seen as a nominal scheme \eqref{eqn_eDDPC} with bounded input disturbance. Such a scheme is recursively feasible and practically stable. Further, we have previously shown in Section~\ref{appendix_guarantees1} that the difference between the optimal input of \eqref{eqn_eDDPC} and that of \eqref{eqn_robust_eDDPC} at time $t$ is bounded by a $\mathcal{K}_{\infty}$ function which depends on the noise level (see \eqref{eqn_bndoptinputs}). Therefore, one can use Lyapunov arguments to show stability of the $n$-step robust eDDPC scheme \eqref{eqn_robust_eDDPC}. Specifically, \eqref{eqn_LyapBnd1} still holds (which is \eqref{eqn_LyapBnds}) as in the previous section. The decay condition \eqref{eqn_LyapDecay} can be obtained by combining \eqref{eqn_bndoptinputs} and \eqref{eqn_LyapDecay4}, with $\phi\coloneqq\phi_4\circ\phi_3\in\mathcal{K}_{\infty}$. Finally, for sufficiently small $\varepsilon^*$, and due to $V(\xi_t^{\mathrm{on}}) \leq \overline{V}$, inequality \eqref{eqn_LyapDecay4} can be further bounded for all $\bar\varepsilon\leq\varepsilon^*$ by $V(\xi_{t+n}^{\mathrm{on}}) \leq \overline{V}$. Therefore, by repeatedly applying all the above arguments, \eqref{eqn_LyapDecay} holds for $t=in$, $i\in\mathbb{N}$. \hfill$\blacksquare$

\end{document}